\documentclass[letterpaper,twocolumn,10pt]{article}
\usepackage{usenix}

\usepackage{tikz}
\usepackage{amsmath}
\usepackage{enumitem}
\setlist[itemize]{nosep,leftmargin=*}

\usepackage{filecontents}
\usepackage{amsmath,amssymb}
\usepackage{booktabs}
\usepackage{rotating}
\usepackage{xurl}

\begin{document}

\date{}

\title{\Large \bf LLMscope: Extracting LLM Assets from Edge AI Chips via Optical Probing
}

\author{
{\rm Dev Mehta, Lily Dukette, William Folan, Olivia Kochol, Noah Solomon, Shahin Tajik, Fatemeh Ganji}\\
Worcester Polytechnic Institute
} 

\maketitle

\begin{abstract}
The move of LLM inference to edge AI accelerators introduces new physical vulnerabilities. 
During execution, model parameters and intermediate inference states are repeatedly loaded into and processed on the chip, making them susceptible to physical side-channel attacks. 
In this work, by deploying laser voltage imaging, we show that one can extract LLM assets during inference, namely embeddings, attention, and quantized MLP weights, activations, and other inference states, from localized memories and compute subcircuits. 
To validate our claims, we perform an attack on an FPGA-based LLM accelerator. 
Since such accelerators reuse the same buffers and compute subcircuits across addresses, tiles, modules, and layers, reading asset values comes down to probing different memories during inference. 
We demonstrate full recovery of the targeted values; however, we also establish a methodology to recover asset values even if some weights or bits remain unread. 
We further derive lower bounds that relate imaging effort to asset dimensions and show that even direct recovery scales linearly with the size of the targeted asset. 

\end{abstract}

\section{Introduction}\label{Sec:Introduction}

Large language model (LLM) inference is moving beyond centralized cloud infrastructure toward hardware deployed close to users.
Saad-Falcon et al. evaluate more than 20 local language models, eight hardware accelerators, and one million real-world single-turn chat and reasoning queries, finding that 88.7\% of the studied queries can be handled by at least one local model and that the coverage of state-of-the-art local models increased from 23.2\% in 2023 to 71.3\% in 2025~\cite{saadfalcon2025intelligence}. 
This shift is accelerated by rapid reductions in the hardware footprint of capable LLM inference. 
Recent measurements show that a workstation costing approximately CHF~8,500 and operating from a standard power outlet can run open models at roughly GPT-5.1-level performance for common workloads, while even a 284B-parameter MoE model can be executed locally for latency-tolerant tasks~\cite{souverana2026selfhosted}.
LLM assets that were once largely confined to physically controlled data-center infrastructure can therefore increasingly reside on compact, locally deployed hardware within direct physical reach of an adversary. 
Protecting the assets processed on such devices is therefore necessary to preserve the confidentiality objective that motivates self-hosted inference.
This transition changes the hardware-security boundary because model parameters and user-dependent inference state are increasingly processed on devices that may be physically accessible to an adversary.
Prior work has already identified on-device model extraction as a practical threat and shown that protecting LLM weights during execution remains difficult even with trusted-execution mechanisms~\cite{nayan2024ondevice,wang2025gameofarrows}.
Physical side-channel attacks against neural-network accelerators further demonstrate that implementation leakage can threaten both model intellectual property and user data~\cite{horvath2024physicalnn}.

Field-programmable gate arrays (FPGAs) are important in this setting because their reconfigurability supports model-specific dataflows, low-precision arithmetic, customized memory hierarchies, and deeply pipelined execution.
These properties are particularly relevant to decoder-only inference, where autoregressive decode repeatedly accesses model weights and key-value (KV) cache entries and is often constrained by memory movement.
FPGA-based systems have consequently been studied for low-batch, energy-constrained, and memory-bound LLM inference, including heterogeneous systems in which graphics processing units (GPUs) execute compute-intensive prefill while FPGAs accelerate decode~\cite{bentoml2023llmhandbook,altera2023aiinference,elastixai2026fpga,achronix2024speedster,chen2024spatialllm,zeng2024flightllm}.
Recent FPGA LLM accelerators differ in scale and organization but share the execution pattern that matters for security: large parameters reside in external memory and are repeatedly transferred into block RAM (BRAM), UltraRAM (URAM), buffers, registers, and compute inputs as inference proceeds~\cite{haris2024secdallm,li2025hummingbird,zeng2024flightllm,liu2025pushinglimits,xu2024llamaf,he2024hlstransform,he2026lutllm,huang2024edgellm,hur2023flexrun,chen2024spatialllm}.
Representative designs stream quantized weights, stage embeddings and KV-cache values, and reuse localized matrix or lookup engines across layers~\cite{li2025hummingbird,liu2025pushinglimits,zeng2024flightllm,xu2024llamaf,he2024hlstransform,he2026lutllm}.
The confidentiality boundary therefore extends beyond external model storage to the model and inference-state assets that repeatedly occupy on-chip memories, registers, and datapaths during execution.

\emph{Model assets} in risk include persistent quantities that define the deployed model, e.g., embeddings, attention and multilayer-perceptron projections, normalization parameters, the vocabulary projection, and the metadata required to interpret quantized, sparse, or lookup-based representations.
On the other hand \emph{inference-state assets} relate to execution-dependent values such as attention state, intermediate hidden vectors, partial sums, KV-cache entries, and logits.
Once these values enter the FPGA fabric, protecting external model storage no longer protects them.
Model-asset exposure threatens model intellectual property, while inference-state exposure can reveal user-dependent computation and provide exact inputs or outputs for recovering downstream model parameters. 
These threats are difficult to assess with conventional power or electromagnetic side channels because LLM inference superimposes activity from many simultaneously active memory and compute structures, obscuring the contribution of individual assets. 
Physical side-channel research has also investigated neural-network extraction through power, electromagnetic, timing, cache, and other implementation-dependent leakage channels~\cite{horvath2024physicalnn}.
However, these works do not establish direct bit-level optical recovery of the model and inference-state assets repeatedly staged by FPGA-based LLM inference. 

\noindent\textbf{Our approach. }We investigate this gap by considering an adversary with physical access to the FPGA, the ability to execute or replay inference. 
Concretely, optical probing is the focus of our study, where the attack requires neither electrical contact with internal signals, modification of the deployed model, nor plaintext access to external model storage. 
We apply electro-optical frequency mapping (EOFM), an active optical contactless-probing technique in which a near-infrared (NIR) laser scans the backside of an operating chip, and local electrical activity modulates the reflected optical signal. 
Frequency-selective processing reconstructs a spatial map of circuit regions containing activity at the selected frequency~\cite{monfared2024laserescape,lohrke2016noplace,tajik2017power}.
This spatial selectivity allows EOFM to isolate localized storage and datapath structures that repeatedly carry model-related values during FPGA LLM inference.  

Prior optical contactless-probing work has demonstrated that backside optical access can expose secret-bearing state and internal functionality on FPGAs, including values involved in protecting encrypted FPGA bitstreams~\cite{lohrke2016noplace,tajik2017power,monfared2024laserescape}. 
Our experiments demonstrate that EOFM provides direct digital-value recovery rather than only activity localization.
After localizing a storage or datapath structure, we identify the physical bit positions carrying a targeted value and recover its complete binary representation under controlled execution.
This establishes a bit-level readout for both model and inference-state assets. 
Repeated inference then exposes different addresses, streamed words, tiles, modules, and layers as they traverse the same localized resources.

Direct readout can also be extended when complete optical coverage is unavailable.
For a linear module, directly recovered entries of assets can be removed from the observed relation and the remaining complete entries are recovered by Gaussian elimination from sufficiently independent input-output pairs. 
When only a small number of individual bits remain unresolved, exact downstream states reject candidate completions that are inconsistent with the known computation.
These exact EOFM observations therefore act as anchors across the deterministic inference graph, while hardware reuse amortizes localization across the assets that successively occupy the same physical structures.
We further derive implementation-aware bounds that relate imaging effort to asset size, the amount of information recovered together per replay state, and the additional observations required for hybrid and downstream-constrained recovery.

\begin{figure}[t]
\centering
\includegraphics[width=\columnwidth]{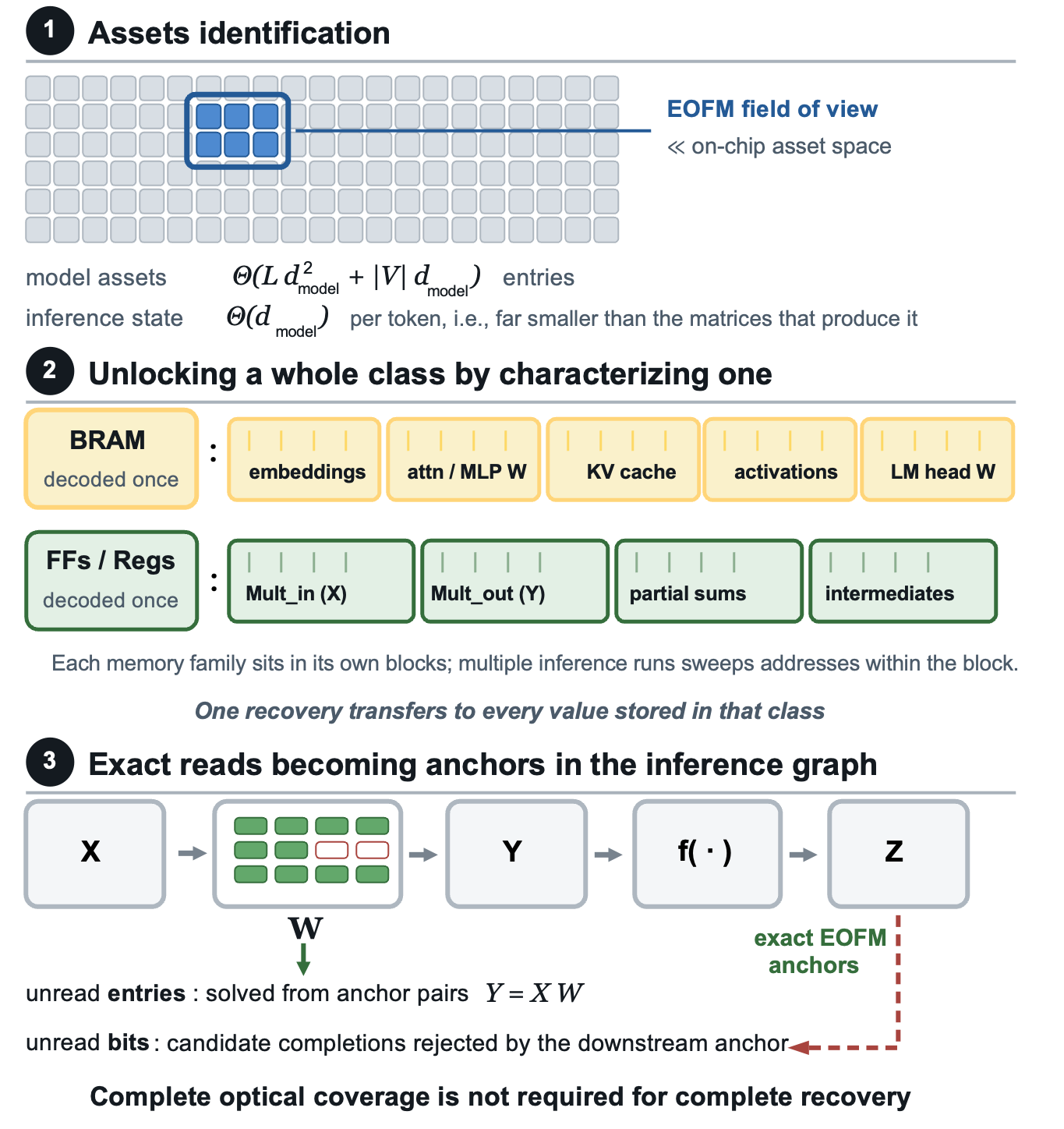}
\caption{Model and inference-state assets are repeatedly staged in the same classes of memory and compute structures.
Direct bit-level reads provide exact anchors, while missing numerical entries or individual bits can be completed using linear input-output relations or recovered downstream states.
Here, $L$ denotes the number of transformer layers, $d_{\mathrm{model}}$ the hidden dimension, and $|V|$ the vocabulary size; $\Theta(\cdot)$ denotes asymptotic bound. 
$X$ is the module input, $W$ the model weights, $Y$ the module output, $f(\cdot)$ the known downstream computation, and $Z$ a later state derived from $Y$.}
\label{fig:LLMscope}
\end{figure}

\noindent\textbf{Contributions in short.}
We make the following contributions:
\begin{itemize}
    \item \textbf{Asset-level security model for FPGA LLM inference.}
    We characterize representative FPGA LLM accelerators as asset-flow systems and identify the on-chip storage, staging, and compute boundaries through which model assets and inference-state assets repeatedly pass.

    \item \textbf{Complete bit-level asset readout with EOFM.}
    We experimentally demonstrate that EOFM localizes FPGA structures carrying LLM-related values and recovers the complete binary representation of targeted assets, advancing optical analysis from activity localization to direct digital-value extraction; see Figure~\ref{fig:LLMscope}. 

    \item \textbf{Recovery beyond complete direct optical coverage.}
    We show how partially recovered linear model assets are completed from EOFM-recovered input-output pairs through Gaussian elimination and how unresolved individual bits are recovered using exact downstream states. 

    \item \textbf{Implementation-aware scalability analysis.}
    We derive imaging-cost bounds that account for asset dimensions, simultaneous bit coverage, replay, and physical resource reuse, and relate these bounds to BRAM, register, stream-width, and buffering characteristics reported by real FPGA LLM implementations. 
\end{itemize}

\section{Foundations and Asset-Level Security Model}
\label{sec:foundations_asset_level_eofm}

We model field-programmable gate array (FPGA) large language model (LLM) inference as an asset-flow problem rather than as a monolithic model implementation.
This abstraction follows the structure of recent FPGA LLM accelerators, in which model parameters and inference state are repeatedly moved from external memory to localized on-chip storage and compute structures before the next token is produced~\cite{haris2024secdallm,li2025hummingbird,zeng2024flightllm,liu2025pushinglimits,xu2024llamaf,he2024hlstransform,he2026lutllm,hur2023flexrun,huang2024edgellm,chen2024spatialllm}.
The security question is not only where the model is stored, but which sensitive values become physically observable during inference.

\subsection{Transformer Modules and Sensitive Assets}
\label{subsec:model_structure_asset_spaces}

We consider a decoder-only transformer and introduce only the modules needed by the security analysis.
Let $d_{\text{model}}$ denote the hidden dimension, $d_{\text{ff}}$ the feed-forward dimension, $n_q$ the number of query heads, $n_{\text{kv}}$ the number of key/value heads, and $d_{\text{head}}$ the head dimension~\cite{Vaswani2017Attention,touvron2023llama2,xu2024llamaf}.
An input token indexes the embedding table $E$, producing a hidden vector that is processed by a sequence of transformer layers.
At layer $\ell$, the attention projections are $Q=XW_Q^{(\ell)}$, $K=XW_K^{(\ell)}$, and $V=XW_V^{(\ell)}$, where $X$ is the layer input and $W_Q^{(\ell)}$, $W_K^{(\ell)}$, and $W_V^{(\ell)}$ are learned model parameters.
The attention block forms token-dependent scores and a context vector, which is mapped back to the hidden dimension through the learned output projection $W_O^{(\ell)}$.
During autoregressive decoding, previously generated key and value vectors are retained in the key-value (KV) cache and reused for subsequent tokens.
The multilayer perceptron (MLP) contains the learned matrices $W_1^{(\ell)}$ and $W_2^{(\ell)}$, and gated architectures additionally contain $W_{\text{gate}}^{(\ell)}$.
Residual and normalization paths operate on the resulting hidden vectors, and the final vocabulary projection $W_{\text{vocab}}$ maps the last hidden representation to logits~\cite{touvron2023llama2,xu2024llamaf}.

Throughout this paper, an \emph{asset} is any confidentiality-sensitive digital value that exists during inference.
We distinguish two classes because their security consequences differ.
\emph{Model assets} are persistent quantities that define the deployed model, including $E$, $W_Q/W_K/W_V$, $W_O$, $W_1/W_2/W_{\text{gate}}$, normalization parameters, $W_{\text{vocab}}$, and the quantization, packing, sparsity, or lookup metadata required to interpret them~\cite{xu2024llamaf,zeng2024flightllm,he2026lutllm}.
\emph{Inference-state assets} are values generated for the current execution, including $Q$, $K$, $V$, attention scores and probabilities, context vectors, partial sums, intermediate hidden vectors, KV-cache entries, and logits.
Recovering model assets threatens model confidentiality and can enable reconstruction of the deployed model, consistent with prior analyses of on-device model extraction and physical side channels against neural-network accelerators~\cite{nayan2024ondevice,horvath2024physicalnn}.
Recovering inference-state assets is independently security-relevant because these values depend on the current prompt, generated context, and execution state; recent attacks have shown that local-LLM side channels and direct KV-cache exposure can reveal or reconstruct user inputs and outputs~\cite{gao2025localcache,luo2026shadowkv}.
Inference-state assets can also expose the exact inputs or outputs of subsequent linear modules and therefore become useful in hybrid parameter recovery.

\begin{figure}[t]
\centering
\includegraphics[width=0.7\columnwidth]{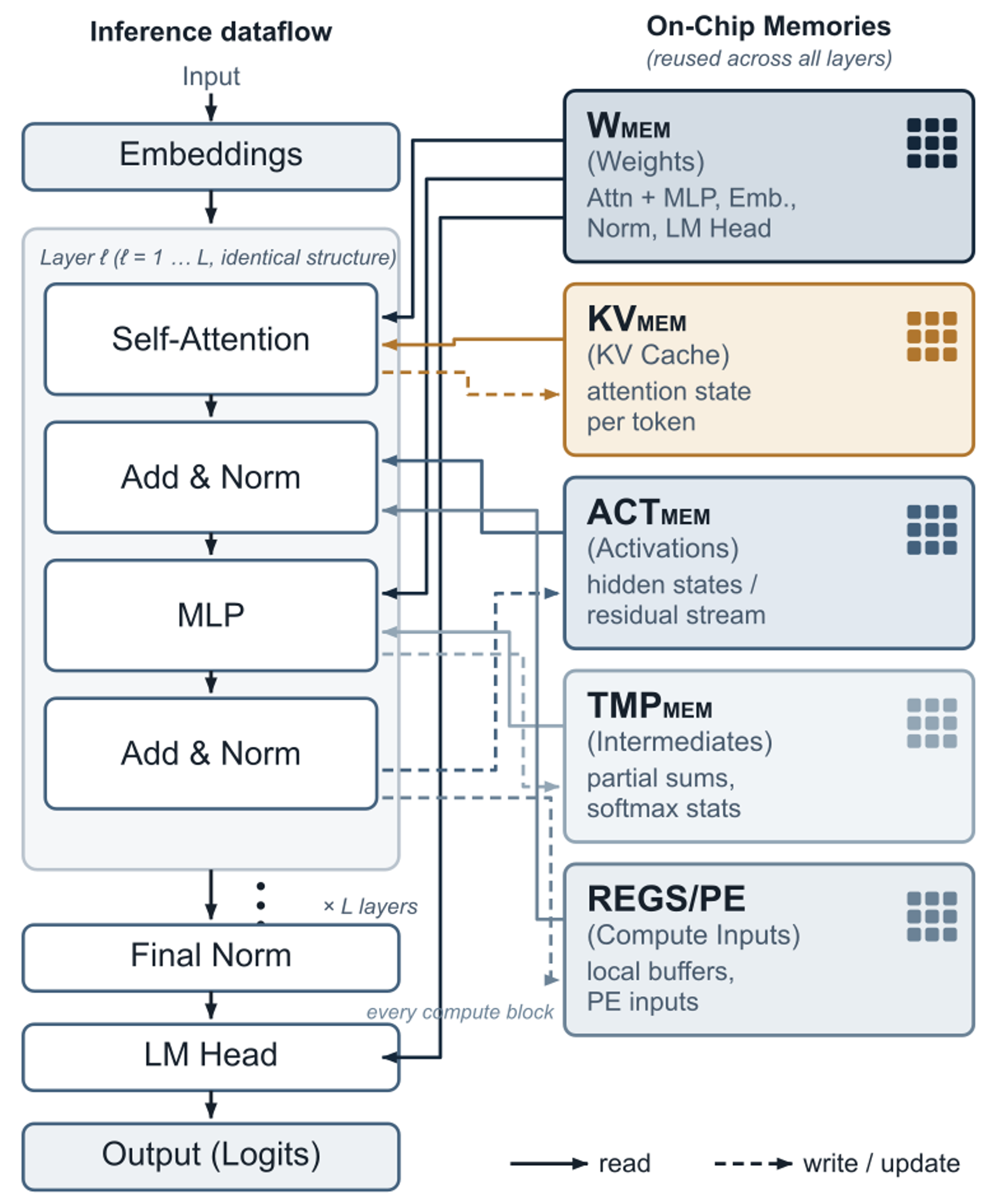}
\caption{Asset flow and on-chip storage during FPGA LLM inference. 
Model parameters and inference-state assets are repeatedly read from and written to reused memory and compute structures as inference proceeds across layers.}
\label{fig:LLM_mem}
\end{figure}
\subsection{Asset Flow in FPGA LLM Inference}
\label{subsec:fpga_llm_asset_flow}

Figure~\ref{fig:LLM_mem} summarizes the common asset flow in FPGA LLM accelerators.
Large model parameters are commonly stored in double-data-rate (DDR) memory, high-bandwidth memory (HBM), graphics double-data-rate (GDDR) memory, or host-accessible memory and are transferred into block random-access memory (BRAM), UltraRAM (URAM), static random-access memory (SRAM), first-in first-out (FIFO) buffers, register files, and processing-element registers as needed~\cite{zeng2024flightllm,xu2024llamaf,li2025hummingbird,liu2025pushinglimits,achronix2024speedster}.
Although the precise hierarchy changes across platforms, the security-relevant sequence is consistent: an asset is stored, transferred, staged near the computation, consumed by arithmetic or lookup logic, and written to another register or buffer.

Representative implementations instantiate this flow in different ways.
LlamaF transfers quantized layer parameters and scale factors from off-chip storage, FlightLLM stages weights, activations, and KV-cache data across HBM, BRAM, and URAM, Hummingbird moves weights, KV-cache values, scales, and intermediate activations through localized memory and compute engines, and Pushing up to the Limit repeatedly streams quantized weights while retaining intermediate hidden state on chip~\cite{xu2024llamaf,zeng2024flightllm,li2025hummingbird,liu2025pushinglimits}.
Spatial, high-level-synthesis, heterogeneous, and lookup-based accelerators change the organization but retain the same security-relevant property that model-derived and input-dependent values occupy localized hardware resources during inference~\cite{chen2024spatialllm,he2024hlstransform,huang2024edgellm,he2026lutllm}.

The relevant protection boundary is therefore the \emph{on-chip lifetime} of an asset.
Protecting model parameters only while they are stored in external memory does not protect the same values after they have been converted to the representation required by the accelerator and placed in an on-chip buffer, register, lookup structure, or datapath.
The same issue applies to transient inference state, including KV-cache windows and intermediate hidden vectors, even when those values exist on chip only for a short interval.
This execution-time exposure is the basis of the asset-level security assessment that follows.

\subsection{EOFM Observation and Adversary Model}
\label{subsec:eofm_visibility_model}

Electro-optical frequency mapping (EOFM) is an active, backside, contactless optical technique developed for integrated-circuit failure analysis~\cite{beyreuther2020eofm,liu2022eofmvoltage}.
A near-infrared laser is scanned across an operating device, local electrical activity modulates the reflected optical signal, and frequency-domain processing produces a spatial map of regions whose activity contains a selected frequency component~\cite{beyreuther2020eofm,liu2022eofmvoltage}.
Prior security work has shown that electro-optical probing and EOFM can be repurposed to infer security-sensitive internal state, including key-dependent behavior in logic-locked circuits~\cite{zuzak2022clap,wojtal2024eofmmitigation}.
Backside optical probing has also been used to expose secret-bearing state on FPGAs and has motivated FPGA-compatible detection and mitigation techniques~\cite{tajik2017power,lohrke2016noplace,monfared2024laserescape}.

Our experiments establish a stronger primitive than activity localization.
After localizing the relevant FPGA structure, EOFM can identify the physical bit positions carrying a targeted value and recover the complete binary representation of that value under controlled execution.
We therefore call an asset \emph{EOFM-readable} when the bits needed to reconstruct its deployed representation can be recovered while that asset is present in an observable on-chip structure.
The definition applies equally to a model asset such as a quantized weight word and to an inference-state asset such as an activation or KV-cache entry.
We consider an adversary with physical backside access to the FPGA, the ability to execute or replay inference, and access to the optical equipment required for EOFM.
The adversary does not require electrical contact with internal signals, modification of the deployed model, or plaintext access to the external model store, and the device remains operational while measurements are collected.
Controlled replay fixes the bitstream, placement and routing, clock and reset schedule, token sequence, quantization configuration, relevant pipeline state, and relevant KV-cache state; stochastic sampling is disabled, fixed, or placed outside the measurement window when it would otherwise change the targeted state.
The attacker can therefore reuse a localized hardware boundary while targeting different spatial regions, pipeline stages, memory banks, addresses, streamed tiles, modules, or layers.
This enables block-by-block recovery even when the full model never resides on chip at once, as in FPGA LLM designs that sequentially load or stream layer parameters through reused matrix engines and buffers~\cite{bentoml2023llmhandbook,altera2023aiinference,zeng2024flightllm,chen2024spatialllm,xu2024llamaf,liu2025pushinglimits}.

\subsection{Direct vs. Indirect Asset Recovery}
\label{subsec:hybrid_direct_algebraic}

Direct bit-level readout is the primary recovery mechanism in this work.
If all bits of a targeted model or inference-state asset are observed as that value passes through an EOFM-readable structure, the deployed representation can be reconstructed without inferring it from aggregate leakage.
For large model matrices, however, full spatial and temporal coverage can require many measurements because FPGA LLM accelerators commonly stream or sequentially load weight blocks rather than keeping the complete model on chip~\cite{xu2024llamaf,zeng2024flightllm,liu2025pushinglimits,li2025hummingbird}.
In that case, direct readout can be combined with the linear relations already implemented by transformer modules.

Consider a linear operation $Y=XW$,
where $W\in\mathbb{R}^{d_{\mathrm{in}}\times d_{\mathrm{out}}}$ is a model asset, $d_{\mathrm{in}}$ and $d_{\mathrm{out}}$ are its input and output dimensions, $X$ contains observed module inputs, and $Y$ contains the corresponding outputs.
This relation applies directly to the query, key, and value projections, the attention output projection, the linear MLP projections, and the vocabulary projection before any subsequent nonlinear operation~\cite{Vaswani2017Attention,zeng2024flightllm,xu2024llamaf}.
Suppose EOFM directly recovers complete numerical values for some entries of a column $w_j$ of $W$, while the remaining entries are unknown.
Partition the corresponding columns of $X$ into $X_{\mathrm{known}}$ and $X_{\mathrm{unknown}}$ so that $y_j=X_{\mathrm{known}}w_j^{\mathrm{known}}+X_{\mathrm{unknown}}w_j^{\mathrm{unknown}}$.
The contribution of the directly recovered entries can be removed,
\begin{equation}
\widetilde{y}_j
=
y_j-X_{\mathrm{known}}w_j^{\mathrm{known}}
=
X_{\mathrm{unknown}}w_j^{\mathrm{unknown}} .
\label{eq:partial_asset_solve}
\end{equation}
If $u_j$ entries remain unknown in column $j$, then $m\geq u_j$ exact input-output observations are sufficient in the noiseless case when $X_{\mathrm{unknown}}$ has full column rank.
The remaining entries can then be obtained by Gaussian elimination or an equivalent linear solve.
Thus, direct EOFM recovery reduces the algebraic problem from $d_{\text{in}}$ unknowns per output column to only the entries that were not read directly.

Hybrid recovery applies to missing \emph{complete numerical entries}, rows, columns, or tiles.
If only selected bits of numerical weight elements are missing, ordinary Gaussian elimination does not directly solve the remaining bit constraints.
Quantized implementations must also account for scales, zero points, group boundaries, packing rules, and rounding whenever these affect the relation between the stored bits and $XW$~\cite{xu2024llamaf,zeng2024flightllm,liu2025pushinglimits}.

To compare the optical costs of direct and hybrid recovery, let $N(Z)$ denote the number of EOFM images required to recover quantity $Z$ under the chosen measurement procedure. 
If $W_{\mathrm{known}}$ is the portion of $W$ recovered directly and one exact $(X,Y)$ observation costs $N(X,Y)$ images, then the hybrid route is advantageous whenever
\begin{equation}
N(W_{\mathrm{known}})+mN(X,Y) < N(W),
\qquad
m\geq \max_j u_j .
\label{eq:hybrid_cost}
\end{equation}
The right-hand side is the cost of directly covering the complete weight matrix, while the left-hand side combines direct recovery of the readable portion with enough exact module input-output pairs to solve the remainder.
Hybrid recovery is advantageous when this combined cost is lower than completing direct weight coverage, which is particularly relevant when a streamed weight matrix spans many temporal states while its input and output vectors appear in smaller, repeatedly reused buffers~\cite{xu2024llamaf,zeng2024flightllm,liu2025pushinglimits}.
Its cost depends on the recovered fraction of $W$, the imaging cost of the $X/Y$ boundaries, the rank of the observed inputs, and the arithmetic representation implemented by the accelerator.

\noindent\textbf{Downstream consistency recovery.}
The sequential structure of LLM inference provides an additional recovery mechanism when the missing information consists of individual bits rather than complete weight entries.
If a small number of bits in an upstream asset cannot be read directly, but the intervening computation and an exact later state are known, candidate completions can be propagated through the known computation and checked against the EOFM-recovered downstream state~\cite{xu2024llamaf,zeng2024flightllm}.
If $k$ bits are missing, at most $2^k$ candidate completions remain; candidates inconsistent with the recovered downstream state are eliminated across one or more controlled inputs until a single completion remains.
This \emph{downstream consistency recovery} does not require analytical inversion, only that the unresolved bits influence a recovered downstream state strongly enough to distinguish the candidates.
Recovered weights or bit completions are validated on held-out controlled inputs by checking consistency with independently EOFM-read module outputs or downstream states. 
Linear and residual boundaries admit direct algebraic constraints, while RoPE is reversible with known parameters. 
Normalization removes offset, scale, or both, while softmax is invariant to a common additive shift; these properties prevent unique inversion from the corresponding output alone.
Nonlinear boundaries can nevertheless participate in candidate checking when the required parameters and intermediate states are known~\cite{su2021roformer,ba2016layernorm,zhang2019rmsnorm,shazeer2020glu,xu2024llamaf,touvron2023llama2}.


If a fraction $\rho$ of the complete entries in each column of a linear model asset is recovered directly, the ideal noiseless full-rank condition becomes $m \geq \left\lceil (1-\rho)d_{\mathrm{in}}\right\rceil$.
For instance, for a 4096-input projection representative of several 7B-scale transformer projections~\cite{touvron2023llama2,liu2025pushinglimits,zeng2024flightllm}, 90\% direct recovery leaves at most 410 unknown entries per column, while 99\% leaves 41.

\section{Asset-Level Security Assessment}
\label{sec:asset_level_security_assessment}

We now apply the model above to the principal model and inference-state assets of decoder-only LLM inference.
For each module, the relevant question is which value becomes materialized at an on-chip boundary, whether EOFM can read it directly, and whether a partially recovered linear model asset can be completed using Eq.~\ref{eq:partial_asset_solve}.


\noindent\textbf{Embedding Table. }The embedding table $E$ is a model asset that maps token identifiers to vectors entering the transformer stack.
A particular execution fetches only the rows selected by the input token sequence, so one inference exposes only a subset of the table.
Hummingbird offloads its large embedding table and transfers selected embedding vectors into the accelerator, while Pushing up to the Limit stores the embedding table together with the model weights and KV-cache state under a tightly constrained external-memory budget~\cite{li2025hummingbird,liu2025pushinglimits}.
The relevant EOFM targets are therefore the buffer, interface, or register boundary at which the selected row becomes available to the transformer.
The security relevance of this boundary is supported by recent local-LLM cache side-channel work showing that token values can be inferred from embedding-access behavior and used to reconstruct victim text~\cite{gao2025localcache}.

Embedding recovery is naturally a direct-readout problem rather than a Gaussian-elimination problem.
If the staged embedding vector is EOFM-readable, the attacker can recover its bits directly and repeat the experiment with different token identifiers to cover additional rows.
If the deployed embedding is quantized or packed, the corresponding representation metadata must also be recovered before the extracted bits can be interpreted as numerical embedding values.
Lookup-based implementations reinforce this point because their codebooks and indices are themselves model assets rather than dense matrices~\cite{he2026lutllm}.


\noindent\textbf{Attention Projection Weights. }The query, key, and value projection matrices are high-value model assets because they are instantiated in every transformer layer and repeatedly consumed during inference~\cite{touvron2023llama2,xu2024llamaf}.
FPGA implementations realize these operations using pipelined matrix-vector units, sparse digital signal processing (DSP) chains, or related matrix-processing structures and repeatedly move quantized weight blocks through on-chip buffers or registers~\cite{xu2024llamaf,zeng2024flightllm,he2024hlstransform,hur2023flexrun}.
LlamaF concatenates projection weights for transfer and applies its group-wise quantized matrix-vector kernel to them, while FlightLLM explicitly loads compressed weights into its on-chip memory hierarchy before matrix processing~\cite{xu2024llamaf,zeng2024flightllm}.

These transfer and staging boundaries support both recovery modes.
If the physical lanes carrying a weight word or tile are EOFM-readable, successive tiles of $W_Q$, $W_K$, and $W_V$ can be recovered directly as they reuse the same hardware.
If some entries or tiles remain unread, the exact hidden vector $X$ and corresponding $Q$, $K$, or $V$ output can provide the input-output pairs needed to solve the missing entries with Eq.~\ref{eq:partial_asset_solve}.
Grouped-query attention changes the dimensions of $W_K$ and $W_V$ but not the exposure mechanism because the corresponding model assets must still be transferred and consumed~\cite{xu2024llamaf,li2025hummingbird}.


\noindent\textbf{Attention State and KV Cache. }The query, key, and value activations, attention scores, normalized attention weights, context vectors, and KV-cache entries are inference-state assets.
Their recovery does not by itself reconstruct model parameters, but it reveals execution state that depends on the current input and generated context.
The KV cache is particularly important because it persists across autoregressive decoding steps and is repeatedly read as each new token is generated~\cite{touvron2023llama2,zeng2024flightllm,li2025hummingbird}.

Existing FPGA LLM designs explicitly expose this state through their memory hierarchy.
Hummingbird buffers quantized KV-cache values as part of its grouped-query-attention dataflow, FlightLLM stores large KV-cache state in HBM while staging the portions required by decode, and Pushing up to the Limit stores a partial KV cache alongside model weights in its constrained DDR memory~\cite{li2025hummingbird,zeng2024flightllm,liu2025pushinglimits}.
FlightLLM also keeps decode activations on chip to reduce external-memory traffic~\cite{zeng2024flightllm}.
These optimizations increase reuse and performance, but they also create repeated physical exposure of user-dependent inference state; direct access to KV-cache contents has been shown to permit reconstruction of sensitive user inputs~\cite{luo2026shadowkv}.

Inference-state recovery has two consequences.
First, exact KV-cache entries, hidden vectors, partial sums, or logits are confidentiality-sensitive even when no model parameter is recovered~\cite{horvath2024physicalnn,gao2025localcache,luo2026shadowkv}.
Second, a recovered inference-state vector can be the exact $X$ or $Y$ required for hybrid recovery of a downstream linear model asset.
We therefore treat inference-state extraction as an independent attack objective and as a possible enabler of incomplete model-asset recovery.

\noindent\textbf{Attention Output Projection. }The attention output projection $W_O$ is a model asset applied to the context vector produced by attention.
The context vector is internally generated rather than directly selected by the adversary, but the FPGA implementation must still materialize it before or during the output projection.
The corresponding weight tiles, context vectors, and projected outputs can therefore appear in FIFO buffers, BRAM/URAM, registers, or processing-element boundaries.

If $W_O$ is completely EOFM-readable at the weight-transfer or staging boundary, it can be reconstructed directly.
If only part of $W_O$ is directly recovered, the EOFM-recovered context vectors and projected outputs provide the $X$ and $Y$ values required by Eq.~\ref{eq:partial_asset_solve}.
This module therefore illustrates why inference-state leakage and model-asset recovery should not be treated as unrelated problems.
The same observable boundary can reveal confidential runtime state and reduce the remaining cost of recovering a persistent model parameter.


\noindent\textbf{MLP Weights. }The MLP is a primary model-extraction target because its expansion and contraction matrices constitute a substantial portion of the per-layer parameter set in Llama-family models~\cite{touvron2023llama2,xu2024llamaf}.
For each layer, the model assets include $W_1$, $W_2$, and, for gated architectures, $W_{\text{gate}}$.
FPGA LLM accelerators implement these operations using matrix-vector or matrix-matrix engines, DSP arrays, sparse DSP chains, pipelined multiplier structures, or mixed-precision compute blocks~\cite{zeng2024flightllm,he2024hlstransform,huang2024edgellm,chen2024spatialllm}.
LlamaF, for example, transfers and processes the MLP projections using the same quantized matrix-vector infrastructure used for other linear layers~\cite{xu2024llamaf}.

Direct EOFM readout of the weight representation is unaffected by the nonlinear activation that follows the first projection.
Hybrid recovery requires more care.
For $W_1$ or $W_{\text{gate}}$, the observed $Y$ must be the linear pre-activation output, because a value observed only after SiLU or another nonlinearity no longer satisfies the simple relation $Y=XW$.
For $W_2$, the activated intermediate vector is the linear input and the down-projection output is the corresponding $Y$ value.
The MLP therefore provides a particularly important use case for the hybrid strategy because its matrices are large, while the input and output vectors can be much smaller than the complete streamed weight representation.


\noindent\textbf{Residual, Normalization, and Elementwise State. }Residual paths and elementwise units manipulate inference-state vectors and smaller model parameters.
FPGA LLM accelerators may implement root-mean-square normalization (RMSNorm), LayerNorm, rotary positional embedding (RoPE), sigmoid linear unit (SiLU), softmax, and related operations in vector or special-function units~\cite{bentoml2023llmhandbook,hur2023flexrun,he2024hlstransform}.
FlightLLM uses a dedicated special-function unit for normalization, activation, softmax, and elementwise operations while moving activations through its memory-management structures~\cite{zeng2024flightllm}.

These modules contain fewer persistent parameters than attention and MLP projections, but they remain relevant.
Normalization scales and related small parameter vectors are required for exact reproduction of model behavior.
The hidden vectors passing through these units are inference-state assets and can also become inputs to later linear modules.
We therefore treat residual and normalization structures as secondary model-asset targets but important inference-state targets.

\noindent\textbf{Vocabulary Projection and Logits. }The final vocabulary projection $W_{\text{vocab}}$ maps the last hidden vector to logits.
The projection matrix is a model asset, while the logits are inference-state assets that determine the distribution used for next-token selection.
LlamaF accelerates the final classifier using the same quantized matrix-vector infrastructure used for other large model matrices~\cite{xu2024llamaf}.

The large vocabulary dimension makes complete direct coverage expensive but does not change the recovery principle.
If $W_{\text{vocab}}$ is streamed through an EOFM-readable weight buffer or register boundary, it can be recovered tile by tile.
If only part of the matrix is directly recovered, exact final hidden vectors and the corresponding logit coordinates provide the input-output pairs needed to solve the missing entries.
Some language models tie the input embedding and output projection weights, so the amount of distinct model state that must be recovered is implementation-dependent~\cite{press2017outputembedding}.
We therefore treat them as separate asset classes unless the evaluated model explicitly shares them.


\noindent\textbf{Quantization, Sparsity, and Lookup Metadata. }Recovering the stored bits of a model parameter is useful only when the deployed representation can be interpreted.
This point is important for FPGA LLM accelerators because low-bit quantization, group-wise scaling, mixed precision, sparsity, and lookup representations are used to reduce memory traffic and fit larger models into constrained devices~\cite{xu2024llamaf,zeng2024flightllm,liu2025pushinglimits,he2026lutllm}.
LlamaF stores quantized weights together with scale factors, while FlightLLM uses compressed and mixed-precision representations together with the metadata required by its mapping flow~\cite{xu2024llamaf,zeng2024flightllm}.
LUT-LLM replaces much of conventional dense multiplication with learned codebooks and quantization indices, making them first-class model assets~\cite{he2026lutllm}.

Accordingly, the extraction target includes both data bits and the metadata required to decode them.
For a quantized matrix, this can include scales, zero points when present, group boundaries, bit widths, and packing order.
For a sparse representation, the nonzero values must be accompanied by the corresponding indices or masks.
For a lookup representation, codebooks, indices, and table organization define the effective model.
This broader asset definition is necessary to distinguish raw bit readout from a usable reconstruction of the deployed model representation.

\subsection{Asset Dimensions and Recovery Scaling}
\label{subsec:resource_aware_image_complexity}
\label{subsec:asset_spaces_recovery_scaling}

The imaging cost of an LLM asset is determined jointly by its deployed size and by how much of that asset is exposed in parallel at the physical boundary being read.
Let $S$ be the number of bits in the deployed representation of the target asset and retain $N(Z)$ as the number of EOFM images required to recover quantity $Z$.
One EOFM image denotes one raster frequency map acquired for one controlled replay condition and one selected frequency.
Let $R$ be the number of images required per logical replay state after accounting for repeated acquisitions or differential conditions, and let $q$ be the number of replay states required to cover the target across addresses, streamed words, time-multiplexed tiles, or physical fields of view.
After one-time localization and calibration, the dominant image count is $N(Z) \approx Rq$. 
For a register or streamed boundary, let $w$ be the number of new target bits recovered together in one replay state.
Complete direct coverage then requires $q\geq\lceil S/w\rceil$, and therefore $N(Z)\geq R\lceil S/w\rceil$.
This relation is the primary direct-recovery bound: larger assets require more coverage, but the cost decreases with the number of asset bits recovered together in each measurement.

\noindent\textbf{Storage organization.}
AMD UltraScale and UltraScale+ devices provide 36-Kb BRAM blocks and 288-Kb URAM blocks, with BRAM supporting widths up to 72 bits in simple-dual-port mode and URAM organized as $4096\times72$~\cite{amdUG573Memory}. 
Let $n_B$ and $n_U$ denote the numbers of BRAM and URAM blocks occupied by an asset, respectively.
If $C_B$ and $C_U$ denote the raw capacities of one BRAM and one URAM block, respectively, then capacity alone gives
$n_B \geq \left\lceil S/C_B\right\rceil$ and
$n_U \geq \left\lceil S/C_U\right\rceil$; storing the same $S$ bits entirely in flip-flops requires at least $S$ data flip-flops~\cite{amdUG573Memory,amdUG574CLB}.
Practical designs can use additional blocks for width, banking, ports, replication, or ping-pong buffering.
Table~\ref{tab:eofm_resource_scale} shows the scale of the register and embedded-memory resources already used by representative FPGA LLM accelerators.

\begin{table*}[t]
\centering
\caption{Published FPGA LLM resource scales relevant to EOFM image planning.}
\label{tab:eofm_resource_scale}
\scriptsize
\begin{tabular}{p{0.16\textwidth} p{0.1\textwidth} p{0.1\textwidth} p{0.1\textwidth} p{0.42\textwidth}}
\toprule
Design & Flip-flops & BRAM & URAM & Security-relevant implementation characteristic \\
\midrule
LlamaF~\cite{xu2024llamaf,li2025hummingbird}
& $\approx171$K
& 223
& --
& TinyLlama-scale embedded inference with quantized layer weights, scale factors, BRAM-cached activations, and pipelined matrix-vector computation. \\
Pushing up to the Limit~\cite{liu2025pushinglimits,li2025hummingbird}
& $\approx105$K
& $\approx37$
& 10
& LLaMA2-7B W4 inference on KV260 with a 4~GB external-memory budget, streamed weights, on-chip hidden state, and partial KV-cache storage. \\
Hummingbird~\cite{li2025hummingbird}
& 25,422
& 59
& 18
& LLaMA3-8B embedded inference with a 512-bit weight bus, 4-bit model weights, 8-bit KV cache, and time-multiplexed on-chip KV-cache buffering. \\
FlightLLM~\cite{zeng2024flightllm}
& 943K
& 1,252
& 792
& U280 implementation with a large HBM-backed model, on-chip decode activations, and a buffer subsystem using 816 BRAMs and all 792 URAMs. \\
\bottomrule
\end{tabular}
\end{table*}

\noindent\textbf{Asset dimensions.}
The transformer architecture determines $S$ for each model or inference-state asset.
We retain the previously defined $d_{\mathrm{model}}$, $d_{\mathrm{ff}}$, $n_q$, $n_{\mathrm{kv}}$, and $d_{\mathrm{head}}$, and let $\mathcal{V}$ denote the vocabulary, $L$ the number of transformer layers, $B$ the batch size, and $T$ the number of tokens represented by the state under consideration.
Using the row-vector convention, the model assets occupy 
\begin{align}
E &\in\mathbb{R}^{|\mathcal{V}|\times d_{\mathrm{model}}}, &
W_Q^{(\ell)} &\in\mathbb{R}^{d_{\mathrm{model}}\times n_qd_{\mathrm{head}}}, \nonumber\\
W_K^{(\ell)},W_V^{(\ell)} &\in\mathbb{R}^{d_{\mathrm{model}}\times n_{\mathrm{kv}}d_{\mathrm{head}}}, &
W_O^{(\ell)} &\in\mathbb{R}^{n_qd_{\mathrm{head}}\times d_{\mathrm{model}}}, \nonumber\\
W_1^{(\ell)},W_{\mathrm{gate}}^{(\ell)} &\in\mathbb{R}^{d_{\mathrm{model}}\times d_{\mathrm{ff}}}, &
W_2^{(\ell)} &\in\mathbb{R}^{d_{\mathrm{ff}}\times d_{\mathrm{model}}}, \nonumber\\
W_{\mathrm{vocab}} &\in\mathbb{R}^{d_{\mathrm{model}}\times|\mathcal{V}|}.
\label{eq:asset_spaces}
\end{align}
The orientation of $W_{\mathrm{vocab}}$ follows the same $Y=XW$ convention; storing its transpose does not change the number of model entries.
For standard multi-head attention, $n_q=n_{\mathrm{kv}}$, whereas grouped-query attention uses $n_{\mathrm{kv}}<n_q$~\cite{ainslie2023gqa}.
In Llama-family configurations, $n_qd_{\mathrm{head}}=d_{\mathrm{model}}$, making $W_Q$ and $W_O$ square while reducing the width of $W_K$ and $W_V$ under grouped-query attention~\cite{touvron2023llama2,xu2024llamaf}.

A hidden representation occupies $\mathbb{R}^{B\times T\times d_{\mathrm{model}}}$; $Q$ occupies $\mathbb{R}^{B\times T\times n_qd_{\mathrm{head}}}$; $K$ and $V$ each occupy $\mathbb{R}^{B\times T\times n_{\mathrm{kv}}d_{\mathrm{head}}}$; the attention context occupies $\mathbb{R}^{B\times T\times n_qd_{\mathrm{head}}}$; the first and gate MLP projections occupy $\mathbb{R}^{B\times T\times d_{\mathrm{ff}}}$; and the second MLP projection returns to $\mathbb{R}^{B\times T\times d_{\mathrm{model}}}$.
The logits occupy $\mathbb{R}^{B\times T\times|\mathcal{V}|}$, a per-layer KV cache storing both keys and values contains $2BTn_{\mathrm{kv}}d_{\mathrm{head}}$ scalar entries, and full-sequence attention scores contain $Bn_qT^2$ entries~\cite{Vaswani2017Attention,ainslie2023gqa}.
The dominant learned matrices in one gated transformer layer contain $2d_{\mathrm{model}}d_{\mathrm{head}}(n_q+n_{\mathrm{kv}})+3d_{\mathrm{model}}d_{\mathrm{ff}}$ scalar entries.
Under $n_qd_{\mathrm{head}}=\Theta(d_{\mathrm{model}})$ and $d_{\mathrm{ff}}=\Theta(d_{\mathrm{model}})$, the dominant model assets of one layer scale as $\Theta(d_{\mathrm{model}}^2)$, while one hidden state per token scales as $\Theta(d_{\mathrm{model}})$.
Across $L$ layers, the dominant matrix assets scale as $\Theta(Ld_{\mathrm{model}}^2)$, while the embedding and vocabulary projections each contain $|\mathcal{V}|d_{\mathrm{model}}$ entries; tied input/output embeddings reduce these two parameter sets to one~\cite{press2017outputembedding}.

\noindent\textbf{Memory-backed assets.}
For a memory-backed asset, $q$ depends on the number of relevant addresses and the number of physical memory banks observed together.
Let $D$ denote the configured depth of each occupied memory bank.
A BRAM36 can be configured as $512\times72$ in simple-dual-port mode or $1024\times36$ in true-dual-port mode, while a URAM has depth $4096$ at 72 bits~\cite{amdUG573Memory}.
If the asset occupies $n_B$ BRAM banks and one EOFM field can resolve the output regions of $v$ banks at once, then a one-address-per-replay strategy requires at least
$q \geq D\left\lceil n_B/v\right\rceil$.
If only a subset of addresses contains the target, $D$ is replaced by the number of relevant addresses. 

Hummingbird provides a concrete KV-cache example.
For LLaMA3-8B, one 4096-token, 8-bit $K$ or $V$ cache with head dimension 128 contains $4{,}194{,}304$ bits, and Hummingbird reports that this state occupies 16 URAMs~\cite{li2025hummingbird}.
If all 16 relevant URAM output regions are resolved in one field of view, complete address coverage requires $q=4096$ replay states and approximately $4096R$ EOFM images after localization.
If only four of the 16 regions are resolved together, the same strategy requires $q=16{,}384$ states and approximately $16{,}384R$ images.
Hummingbird time-multiplexes the $K$ and $V$ storage, so the same localized memory resource can expose the two assets at different points in the implementation schedule~\cite{li2025hummingbird}.

\noindent\textbf{Register and intermediate-state assets.}
A 4096-element FP16 hidden vector contains 65,536 bits.
If represented entirely in flip-flops it requires at least 65,536 data flip-flops, while raw BRAM capacity would require only two BRAM36 blocks; practical banking can use more blocks to provide the required datapath width~\cite{amdUG573Memory}.
\cite{liu2025pushinglimits,metaLlama2Repo} uses a 4096-dimensional LLaMA2-7B hidden state and a 128-value FP16 vector-processing width. 
At this boundary, $w=2048$ bits are exposed per aligned replay state, so complete vector coverage requires only $q\geq32$ states.
To make this concrete, as approximately 80 distinct target bits are recovered per EOFM image in our setup, each 2,048-bit replay state requires at least 26 images, and the 32-state traversal therefore requires at least 832 images.
At five minutes per image, this minimum image count corresponds to about 69 hours of acquisition; our actual acquisition time is less than five minutes per image.  

\noindent\textbf{Streamed-weight assets.}
A $4096\times4096$ projection quantized to 4 bits contains $67{,}108{,}864$ bits, or 8~MiB.
This projection scale is representative of LLaMA2-7B, and embedded FPGA LLM designs in this regime use 4-bit weights and wide memory interfaces~\cite{metaLlama2Repo,liu2025pushinglimits,li2025hummingbird}.
If a fully packed EOFM-readable 512-bit weight path exposes one new word per replay state, complete direct traversal requires at least $q=131{,}072$ states before repetition, additional fields of view, or separate metadata measurements are counted.
This is the direct-coverage lower bound for such a fully packed path; interleaved scales, control information, or distribution across multiple physical regions increase the required coverage~\cite{liu2025pushinglimits,xu2024llamaf,zeng2024flightllm}.
The contrast between 131,072 streamed-weight states and 32 states for the 4096-element activation shows why hybrid recovery can be attractive when remaining weight tiles are costly to image but exact $X/Y$ boundaries are compact and repeatedly available.

\noindent\textbf{Recovery bounds.}
The direct bound $N(Z)\geq R\lceil S/w\rceil$ connects the transformer dimensions above to the physical parallelism of the observed boundary.
For fixed deployed precision, a linear asset with $d_{\mathrm{in}}d_{\mathrm{out}}$ entries has $S=\Theta(d_{\mathrm{in}}d_{\mathrm{out}})$ bits, so direct recovery requires $\Omega(Rd_{\mathrm{in}}d_{\mathrm{out}}/w)$ images.
When these matrices traverse a common reusable boundary exposing $w$ new target bits per replay state, one gated transformer layer requires $\Omega\!\left(\frac{R}{w}[d_{\mathrm{model}}d_{\mathrm{head}}(n_q+n_{\mathrm{kv}})+d_{\mathrm{model}}d_{\mathrm{ff}}]\right)$ images, which becomes $\Omega(Rd_{\mathrm{model}}^2/w)$ under the usual dimension relations.
Across $L$ layers using the same class of boundary, the corresponding model-asset term is $\Omega\!\left(\frac{R}{w}[Ld_{\mathrm{model}}^2+|\mathcal{V}|d_{\mathrm{model}}]\right)$, with a second vocabulary-sized term when $E$ and $W_{\mathrm{vocab}}$ are not tied.
These bounds retain $w$ explicitly because EOFM recovers multiple target bits together, as demonstrated by the FF and BRAM experiments.

For hybrid recovery, if a fraction $\rho$ of the complete entries in each output column is recovered directly, at most $(1-\rho)d_{\mathrm{in}}$ entries remain unknown per column and the full-rank condition requires $m=\Omega((1-\rho)d_{\mathrm{in}})$ independent input-output observations.
The imaging cost of those observations is determined by $N(X,Y)$ at their own physical boundaries, so hybrid recovery directly reduces weight coverage and substitutes exact state measurements when $N(W_{\mathrm{known}})+mN(X,Y)<N(W)$, as established in Section~\ref{subsec:hybrid_direct_algebraic}.
For a square projection and fixed observation width, collecting complete state vectors can retain quadratic worst-case scaling in the matrix dimension, but the practical gain is that compact and reusable $X/Y$ boundaries can replace unread or expensive weight locations.

For downstream consistency recovery, $k$ unresolved bits leave at most $2^k$ candidate completions and therefore require at least $k$ bits of independent downstream discrimination in the ideal case.
The corresponding image cost is set by the amount of useful downstream state recovered together per measurement rather than by $k$ alone.
Because a hidden state scales as $\Theta(d_{\mathrm{model}})$ while a square projection scales as $\Theta(d_{\mathrm{model}}^2)$, a recovered downstream state can provide a substantially smaller consistency anchor when it uniquely distinguishes the remaining candidates.
The offline search can involve up to $2^k$ candidates, trading optical coverage for computation.

\noindent\textbf{Reuse and system-level recovery cost.}
FPGA LLM accelerators repeatedly reuse matrix and buffering resources while different layers and modules traverse them~\cite{xu2024llamaf,zeng2024flightllm}.
Once a buffer or datapath is localized and calibrated, the same physical sites can therefore be reused to read values belonging to different model assets, so increasing model depth increases replayed asset states without requiring proportional growth in independently localized sites.
Let $N_0$ be the one-time image cost of localization and calibration and let $i$ index the linear model assets selected for recovery.
The system-level image count is therefore estimated by
\begin{equation}
N_0
+
\sum_i
\min\!\left[
N(W_i),
N(W_i^{\mathrm{known}})+m_iN(X_i,Y_i)
\right].
\label{eq:system_recovery_cost}
\end{equation}
For each linear asset, the minimum selects complete direct recovery or partial direct recovery followed by algebraic completion, while $N_0$ is amortized across reused hardware.
Downstream consistency provides an additional route when the unresolved information consists of individual bits rather than complete numerical entries.

\section{Experiments and Results}\label{Sec:Experimental Setup}

\noindent\textbf{FPGA Board.}
All experiments were conducted on a Digilent Genesys 2 development board equipped with an AMD/Xilinx Kintex-7 (XC7K325T-2FFG900C) FPGA fabricated on a 28~nm process node~\cite{Genesys2FPGA}.
The device features a flip-chip package, requiring only the removal of the heat spreader and cooling fan to expose the silicon die for backside optical access; no additional mechanical preparation, such as silicon thinning or polishing, was required.
The FPGA was operated at a core supply voltage of 1.0~V and a clock frequency of 200~MHz.
All hardware designs were synthesized, placed, and routed using the Xilinx Vivado Design Suite~\cite{Vivado:2023}.

\noindent\textbf{Optical Setup.}
We used a Hamamatsu PHEMOS-X FA microscope~\cite{hamamatsu:2026} for electro-optical frequency mapping (EOFM).
The system provides backside infrared imaging for navigation and alignment and supports objective lenses with magnifications of 5$\times$/0.14~NA, 20$\times$/0.4~NA, and 50$\times$/0.76~NA, together with additional 2$\times$, 4$\times$, and 8$\times$ optical zoom.
For EOFM, the selected region was raster scanned and the reflected optical signal was measured by the photodetector and processed by the spectrum analyzer to form a frequency-selective spatial map.
The scans were acquired at 0.33~ms/pixel.
Our EOFM measurements used the 50$\times$ objective with 2$\times$ or 4$\times$ zoom depending on the target structure.

\noindent\textbf{DUT.}
The matrix multiplier was adapted from the open-source \texttt{Buck008} project~\cite{Buck008FPGA}, which implements a systolic-array matrix multiplication accelerator.
Systolic architectures are widely used in AI hardware accelerators~\cite{chen2023high,chang2025hardware,zhang2019fault}.
Because our goal is to evaluate the storage and staging structures that carry LLM assets, we removed the original control and communication modules and retained the matrix-multiplication datapath together with BRAM and register boundaries.
The testbed consists of a systolic array operating on signed 8-bit input values and producing signed 17-bit outputs.
BRAMs provide the memory-backed boundary representative of assets such as embeddings, KV-cache entries, activations, and buffered model data, while the input and output registers represent the FF-based staging boundaries used around compute units.
The design operates at 200~MHz and each matrix computation requires 16 clock cycles.

\noindent\textbf{EOFM excitation.}
For both the FF and BRAM experiments, we alternated a target data value with \texttt{0x00} once every 16 clock cycles.
At the 200~MHz system clock, this produces a 12.5~MHz modulation component that allows the physical locations carrying the target bits to be isolated in the EOFM map.
Once the physical bit positions are identified, activity at a position is decoded as logic `1' and the absence of activity as logic `0'.

\subsection{Direct Asset Recovery from FFs}

We evaluate FFs as these registers can carry embeddings, attention and MLP inputs, activations, partial sums, intermediate states, matrix outputs, and streamed or locally buffered weights.
Reading their values therefore evaluates a storage boundary through which both model assets and inference-state assets repeatedly pass during inference.

\begin{figure}[t]
    \centering
    \includegraphics[width=\columnwidth]{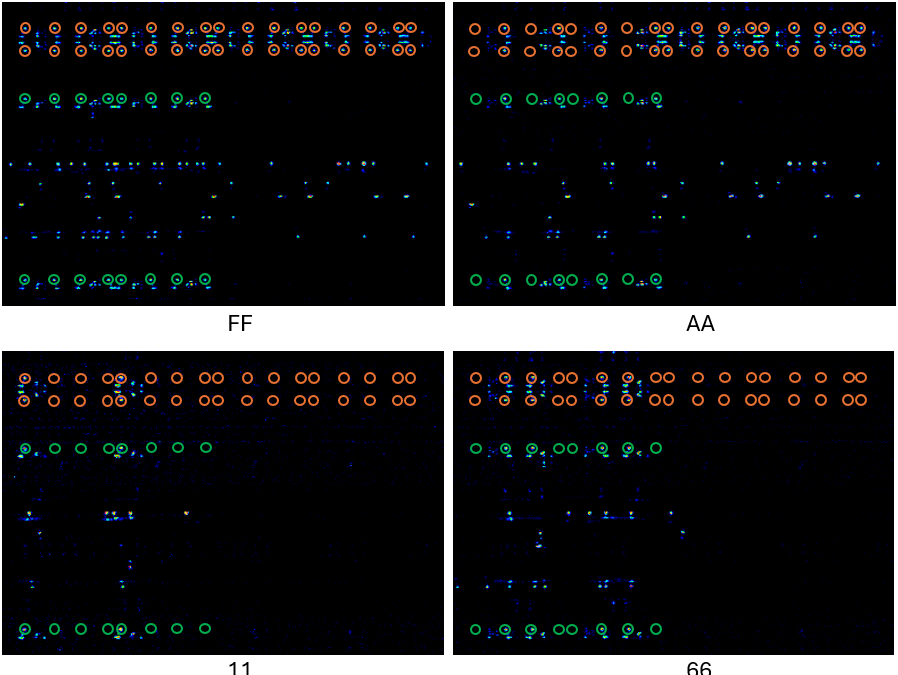}
    \caption{EOFM results for FFs representing the input and output registers of the matrix multiplier.
    Green circles indicate input bits, while orange circles highlight the corresponding output bits for each evaluated input byte.}
    \label{fig:FF_EOFM}
\end{figure}

The FFs in the target Kintex-7 are implemented using FDCE primitives.
We used a tightly placed configuration to represent a dense compute boundary and acquired EOFM maps with the 50$\times$ objective and 2$\times$ zoom at 12.5~MHz.
Figure~\ref{fig:FF_EOFM} shows the input and output register locations for four representative input conditions.
Green circles mark the input bits and orange circles mark the corresponding output bits.
The bit locations are sufficiently resolved to reconstruct the complete input and output representations directly from the EOFM maps.

For example, the \texttt{0x11} panel decodes to two signed 8-bit input values of 17 and two signed 17-bit output values of 17.
The other panels are decoded in the same way.
Across the four measurements, the recovered inputs and outputs match the values produced by the programmed identity matrix, validating the physical bit mapping and numerical decoding.
More importantly, these measurements provide exact input-output states recovered from the FPGA. 
We use these EOFM-decoded states in the recovery experiments below.

\subsection{Direct Asset Recovery from BRAMs}

BRAMs are a primary on-chip storage resource for LLM data in FPGA accelerators.
They can hold embeddings, KV-cache entries, activations, intermediate states, buffered weights, and representation metadata depending on the accelerator.
Demonstrating bit-level recovery from BRAM therefore evaluates the memory-backed counterpart of the FF staging boundary and shows that LLM assets can be read while they reside in the on-chip memory hierarchy.

\begin{figure}[t]
    \centering
    \includegraphics[width=0.85\columnwidth]{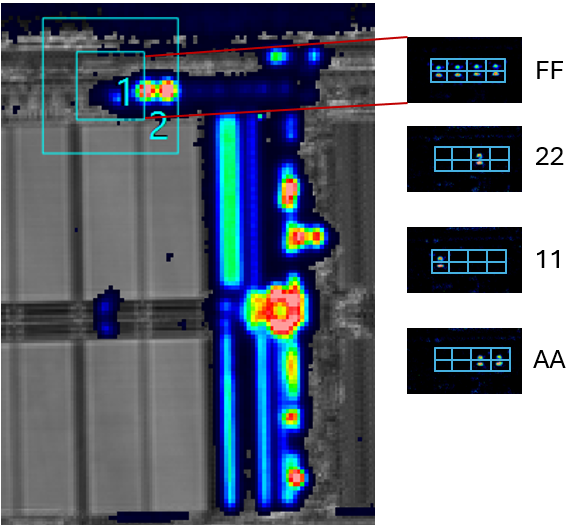}
    \caption{EOFM results for BRAM outputs.
    The left panel provides an overview of the selected BRAM region and the blue box denotes the field of view used for high-magnification EOFM.
    The remaining panels show the EOFM maps for representative data patterns.}
    \label{fig:BRAM_EOFM}
\end{figure}

We first evaluated an 8-bit BRAM configuration with the optional output register disabled and applied the same 12.5~MHz alternating-data pattern.
Figure~\ref{fig:BRAM_EOFM} shows the resulting EOFM maps acquired with the 50$\times$ objective and 4$\times$ zoom.
The data bits form a compact physical pattern that differs from the FF layout.
The \texttt{0xFF} map identifies all active bit positions, while \texttt{0xAA} and \texttt{0x22} reveal the logical-to-physical bit ordering.
In particular, the measurements show that bits 1 and 2 within each 4-bit group are physically exchanged relative to a naive spatial ordering.
Accounting for this mapping allows the complete BRAM output byte to be reconstructed from the EOFM image.

\begin{figure}[t]
    \centering
    \includegraphics[width=0.5\columnwidth]{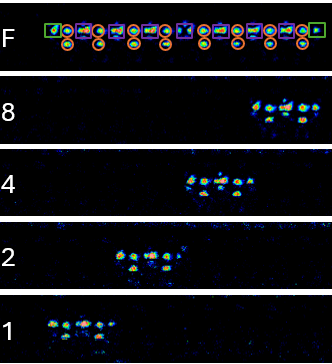}
    \caption{EOFM results for a 16-bit BRAM configured with internal output registers.
    Orange circles denote active bit locations, while bounding boxes mark the physical output register logic.
    Purple boxes enclose 2-bit register pairs, and green boxes delineate single-bit edge registers.
    The 16-bit output is arranged across two 8-bit rows.}
    \label{fig:BRAM_EOFM_16bit}
\end{figure}

We then enabled the BRAM output pipeline registers and increased the data width to 16 bits.
Figure~\ref{fig:BRAM_EOFM_16bit} shows that the wider output remains directly decodable.
The output is arranged across two 8-bit rows, with the physical register boundaries visible in the EOFM maps.
The repeated patterns for \texttt{0x1111}, \texttt{0x2222}, \texttt{0x4444}, and \texttt{0x8888} identify the bit ordering, while \texttt{0xFFFF} confirms all 16 active positions.
Thus, multiple asset bits are recovered together from the same EOFM map rather than requiring a separate image for every bit.

Together, the FF and BRAM experiments establish direct recovery at the two principal storage boundaries evaluated in our testbed.
The recovery procedure depends on the physical structure carrying a value, not on the semantic role of that value in the LLM.
Once a register or BRAM bit mapping is localized, the same readout procedure can recover any LLM asset that is staged in that structure as inference progresses.
The next experiments evaluate how recovery proceeds when complete optical coverage of a larger asset is not available.

\subsection{Recovery Beyond Complete Readout}

The direct experiments above recover complete numerical values from the EOFM images.
As assets scale, however, some weight entries or individual bits may remain unread because the complete asset is not simultaneously visible or because a subset of positions cannot be resolved.
We therefore evaluate three forms of incomplete recovery using only the four FF EOFM captures in Figure~\ref{fig:FF_EOFM}, corresponding to the \texttt{0xFF}, \texttt{0xAA}, \texttt{0x11}, and \texttt{0x66} input conditions.
No additional optical measurements are collected.
Instead, information that was available in the original experiment is deliberately withheld in post-processing so that the recovery mechanisms can be evaluated against known ground truth.

\noindent\textbf{EOFM decoding and recovery procedure.}
The input to the recovery analysis is obtained from the EOFM images rather than from the panel labels.
We automatically crop the four panels and locate the annotated green input circles and orange output circles from their colors.
The colored outline pixels are excluded, the mean image intensity is measured inside each circle, and active and inactive locations are separated using the midpoint of the largest gap in the sorted intensity scores.
Reading bit 0 from the left, the decoded bits are assembled LSB-first and interpreted as two signed 8-bit input values and two signed 17-bit output values per panel.
The panel labels are used only as a post-decoding cross-check.

Missing information is then introduced only after this EOFM decoding step.
For the linear-recovery experiment, we construct the exact system $Y=XW$ from the decoded input-output pairs and use reduced-row-echelon-form Gaussian elimination with rational arithmetic; the matrix rank determines whether $W$ is unique and the nullspace identifies entries that remain unconstrained.
For the bit-level weight experiment, candidate completions of the artificially hidden weight bits are enumerated and propagated through each decoded $X_i$, and a candidate is retained only if its predicted output agrees with every readable bit of the corresponding decoded $Y_i$.
For the missing-input experiment, candidate completions of the hidden input bits are propagated through the known programmed matrix and compared exactly with the decoded downstream output.
Searches are exhaustive where practical; the larger weight/output mask spaces use deterministic seeded sampling with edge cases that include sign, least-significant, and most-significant bit positions.
A recovery is accepted only when a single candidate survives and the recovered value agrees with the unmasked ground truth.

\noindent\textbf{Observation diversity and the rank condition.}
We first ask how many input-output observations is by itself sufficient to determine an otherwise unknown matrix. 
The EOFM maps decode to the four input states, with corresponding 17-bit outputs. 
We evaluate all 15 nonempty subsets of these observations.  
Four subsets with one observation, six with two, four with three, and the complete four-observation set. 
Every stacked input matrix has rank one.
The resulting linear system for the four entries of $W$ has rank two regardless of whether 25\%, 50\%, 75\%, or 100\% of the observations are retained.
The measurements constrain only $w_{00}+w_{10}=1$ and $w_{01}+w_{11}=1$, so none of the four matrix entries is individually unique.
The programmed identity matrix satisfies these constraints although being one of the consistent solutions. 
\begin{figure}[t]
\centering
\includegraphics[width=\columnwidth]{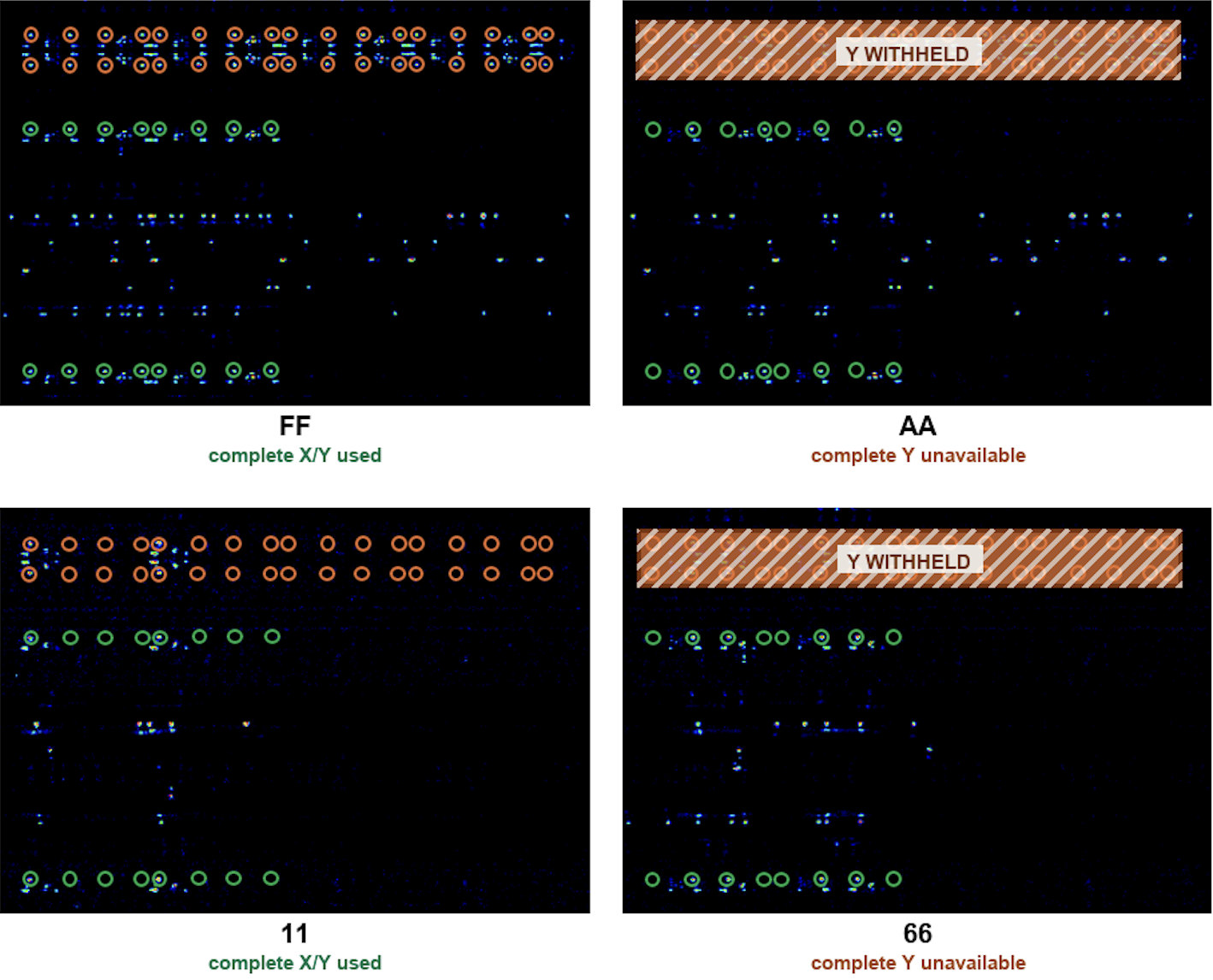}
\caption{Representative 50\% observation availability.
The complete orange output arrays for \texttt{0xAA} and \texttt{0x66} are artificially withheld in post-processing.
All four panels originate from the EOFM captures in Figure~\ref{fig:FF_EOFM}.}
\label{fig:weight_recovery_partial_observations}
\end{figure}
This result directly validates the rank requirement in the hybrid-recovery analysis.
The condition $m\geq u_j$ is sufficient only when the corresponding input submatrix has full column rank.
Additional executions do not improve recoverability when they remain in the same input subspace.
Thus, scaling hybrid recovery requires both a sufficient number of observations and sufficient diversity among those observations.

\noindent\textbf{Joint recovery with incomplete weights and outputs.}
We next evaluate the bit-level case in which both the model asset and its downstream state are incomplete.
The decoded $X$ states remain fully available, while $k_W\in\{1,2,4,8\}$ bits of the 32-bit representation of the $2\times2$ weight matrix and $k_Y\in\{0,1,2,4,8\}$ bits of each 34-bit downstream output state are artificially hidden.
Each assignment to the missing weight bits defines a candidate matrix.
For every recovered input, the candidate is evaluated and rejected whenever a predicted output bit disagrees with a readable output bit.

We evaluate all 24 orders of the four EOFM observations for every selected weight/output mask pair.
Weight masks are exhaustive for $k_W=1$ and $k_W=2$ and deterministically sampled for $k_W=4$ and $k_W=8$, while output masks are exhaustive for $k_Y\leq2$ and deterministically sampled for $k_Y=4$ and $k_Y=8$.
Overall, the experiment evaluates 12,898 weight/output mask pairs and 309,552 observation-order trajectories.
The reported recovery fractions for the larger mask spaces therefore describe the tested deterministic mask sets rather than a probability over all possible masks. 

Recovery remains possible even when both quantities are incomplete.
With one missing weight bit, every tested mask is uniquely recovered for all output-readability levels, including the case in which eight output bits per state are hidden.
With two missing weight bits and eight hidden output bits, 743 of 747 tested mask pairs (99.46\%) uniquely recover the exact programmed matrix.
The corresponding counts are 725 of 764 (94.90\%) for four missing weight bits and 421 of 508 (82.87\%) for eight missing weight bits.
The latter begins with 256 possible weight completions while retaining only 26 of the 34 output bits, or 76.47\% of each downstream state. 

\begin{figure}[t]
\centering
\includegraphics[width=0.7\columnwidth]{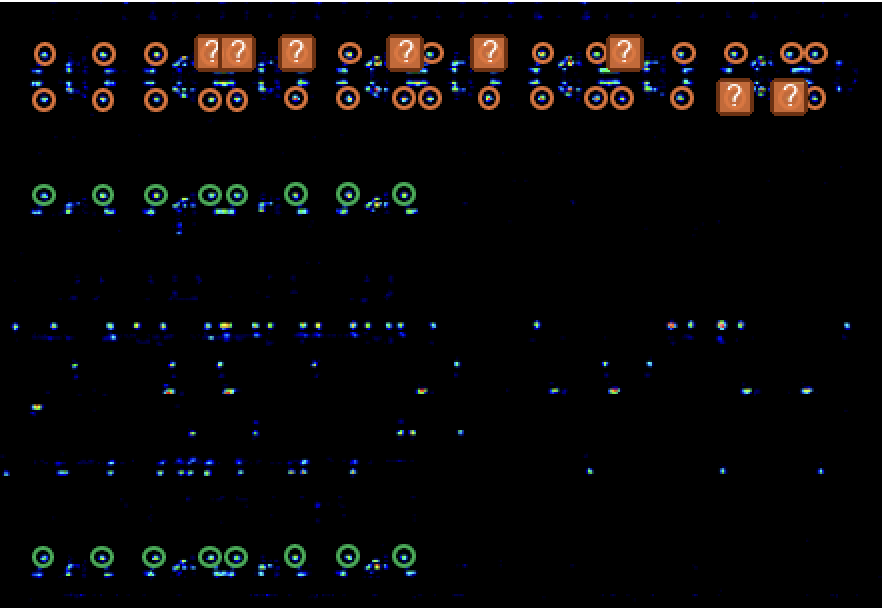}
\caption{Joint recovery when both $W$ and $Y$ are incomplete.
Eight bits of $W$ and eight bits of each downstream state are treated as unreadable.
Applying the EOFM observations in the order \texttt{0xFF}, \texttt{0x11}, \texttt{0x66}, and \texttt{0xAA} reduces the candidate set from $256$ to $4$, $4$, $2$, and finally one candidate.}
\label{fig:joint_incomplete_WY_recovery}
\end{figure}

Figure~\ref{fig:joint_incomplete_WY_recovery} illustrates one of the more demanding successful cases.
Eight unresolved weight bits initially produce $2^8=256$ candidate matrices.
The \texttt{0xFF} observation reduces this set to four.
The \texttt{0x11} observation contributes no additional discrimination, the \texttt{0x66} observation reduces the set to two, and the \texttt{0xAA} observation leaves only the programmed identity matrix. 
Ambiguous cases can also occur.
For example, when the same bit position is missing from both entries contributing to one output coordinate, two candidates can remain even when $Y$ is completely known.
Recovery therefore depends on whether the readable constraints distinguish the candidate completions, not simply on the number of observations collected.
The experiment also shows that the downstream state need not always be bit-complete: a partially recovered downstream state can suffice when its readable bits uniquely discriminate the candidate weights.

\noindent\textbf{Recovering missing upstream bits from downstream state.}
Finally, we evaluate downstream-constrained recovery directly.
Here, the matrix is known and individual bits of the EOFM-decoded upstream state $X$ are artificially treated as unreadable.
We test $k_X\in\{1,2,4,8\}$ missing bits, leaving at most $2^{k_X}$ candidate completions.
Each candidate is propagated through the known matrix operation and compared with the independently EOFM-decoded downstream state $Y$.

We evaluate both 8-bit input rows in all four EOFM panels and exhaustively test every bit-position mask for the selected values of $k_X$.
This gives 64 one-bit masks, 224 two-bit masks, 560 four-bit masks, and eight complete-byte masks, for a total of 856 masking configurations and 12,032 explicitly evaluated candidate completions.
In every case, the downstream state eliminates all incorrect candidates and leaves exactly the original input.
A single downstream observation is sufficient for all 856 masks.
In the strongest case, all eight bits of an input byte are treated as unreadable, producing 256 possible values before the downstream constraint is applied; comparison with the recovered $Y$ shows exact recovery.

\begin{figure}[t]
\centering
\includegraphics[width=0.7\columnwidth]{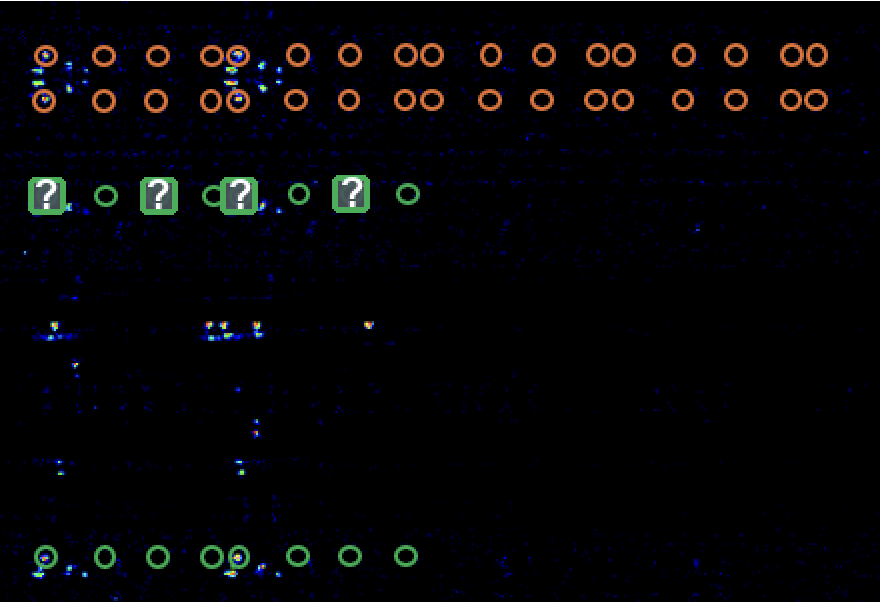}
\caption{Four missing input bits from the representation of \texttt{0x11} are introduced in post-processing.
The complete orange downstream output remains available and uniquely selects the original input completion.}
\label{fig:missing_bit_downstream_recovery}
\end{figure}
\section{Discussion} \label{Sec:Discussion}

\noindent\textbf{Portability across FPGA families and embedded platforms.}
Our experiments use a Kintex-7 device, but the recovery procedure targets the storage structures that carry the asset rather than a device-specific LLM module.
FFs and registers stage values around compute units, while BRAM-like embedded memories buffer model and inference-state data across FPGA accelerator families.
When another FPGA implementation places embeddings, weights, activations, KV-cache entries, or intermediate results in the same class of memory or register structure, the attack follows the same procedure: localize the physical bit positions, replay the target state, and decode the resulting values.
The physical locations and EOFM parameters must be characterized for each FPGA family, but the recovery procedure depends on the storage primitive rather than on the semantic role of the asset.
The same principle extends to embedded accelerators that stage sensitive values in optically accessible register files or SRAM-like memories.

URAM is also relevant in newer FPGA families, it can be a particularly favorable target for EOFM. 
A single URAM primitive stores 288~Kb, eight times the capacity of a 36~Kb BRAM, and URAM resources are organized in dedicated columns on UltraScale+ devices~\cite{amdUG573Memory}. 
Since EOFM provides spatially resolved activity maps that allow active circuit regions to be localized on the die~\cite{tajik2017power}, the larger and readily identifiable URAM structures can simplify localization and optical targeting compared with smaller BRAM resources.

The same storage-centric view also explains why AXI does not require separate treatment.
AXI transports data into the programmable logic, but the received value must ultimately be written into registers, FFs, BRAM, URAM, or another local memory before it is consumed by the computation.
An AXI-loaded model asset therefore reaches the same storage boundaries evaluated in our experiments.
The attack targets the physical structure in which the digital value is staged or stored, not the protocol that delivered it.
Our testbed consequently captures the memory and register boundaries through which model and inference-state assets repeatedly pass in FPGA LLM accelerators.

\begin{figure}[t]
    \centering
    \includegraphics[width=\columnwidth]{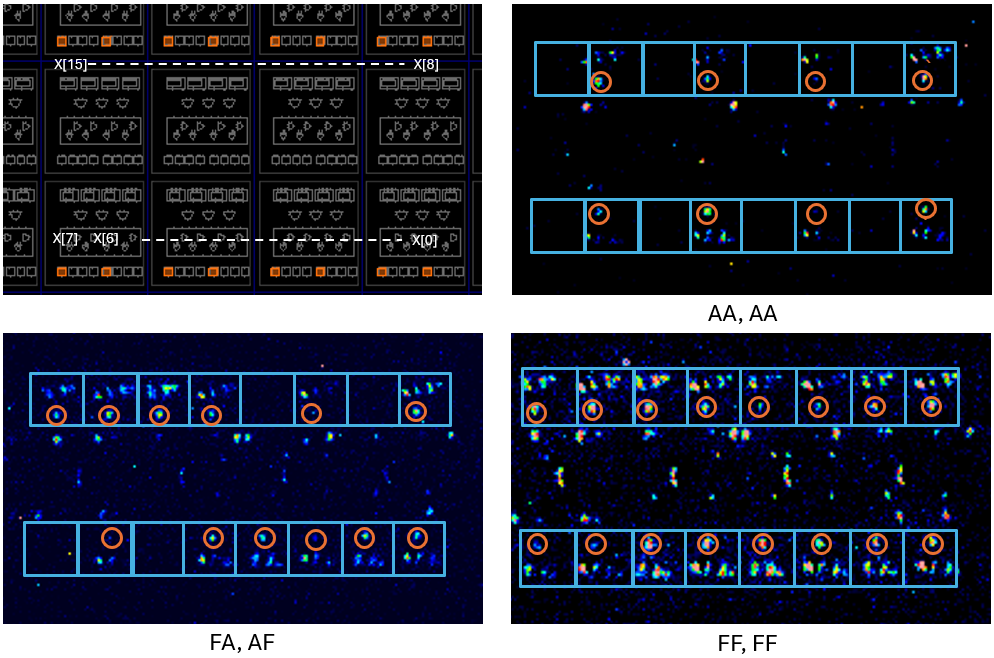}
    \caption{Photon-emission results for sparsely placed FFs forming the matrix-multiplier input registers.
    The top-left panel shows the Vivado floorplan, while the remaining panels show representative data values.
    Blue boxes mark FF locations and orange circles mark active logic `1' states.}
    \label{fig:FF_PE_AN}
\end{figure}

\noindent\textbf{EOFM vs. its alternative, photon emission.}
Photon emission (PE) can reveal digital values in sparse FPGA layouts, but its effectiveness degrades in the dense and time-multiplexed structures relevant to LLM inference.
In our sparse-FF experiment, photon emission with a 20$\times$ objective resolved individual FDCE locations and their corresponding logic states, as illustrated in Figure~\ref{fig:FF_PE_AN}. 
With tightly placed FFs, however, emissions from neighboring FFs and routing activity overlapped and individual bit locations could no longer be cleanly separated.
Temporal reuse creates a second limitation because time-integrated photon emission superimposes activity from the different values processed by a reused compute structure during the acquisition window.
These spatial and temporal limitations arise from passive, time-integrated collection and make complete asset extraction increasingly difficult as placement density and hardware reuse increase.
EOFM addresses this regime by actively selecting the imposed modulation frequency and spatially resolving the corresponding bit locations, enabling direct multi-bit recovery from the dense FF and BRAM structures evaluated in our experiments.
Electro-optical probing (EOP) offers a complementary time-domain alternative for observing the switching waveform of one localized primitive, but trades spatial parallelism for single-site temporal resolution~\cite{amini2024comparative}.

%
%
%

\noindent\textbf{Controlled replay.}
Our experiments use an alternating \texttt{0x00} reference to create a strong periodic contrast, but the zero-valued reference is not fundamental to the attack.
Any repeatable transition, including a controlled reset or pipeline initialization, can provide the periodic contrast required for localization and bit decoding as long as the target asset state is reproduced under stable execution conditions.

\noindent\textbf{Implication for larger LLM accelerators.}
The experiments and bounds together show that scaling the attack does not require independently localizing every model bit: a characterized memory or register boundary can be reused as different assets traverse it, while algebraic and downstream constraints recover information that direct EOFM leaves incomplete.

\section{Conclusion}\label{Sec:conclusion}
We showed that FPGA-based LLM inference creates a direct optical path to both model assets and inference-state assets while these values are staged and processed on chip.
Using EOFM, we demonstrated complete bit-level recovery from the FF and BRAM structures that carry weights, embeddings, activations, KV-cache entries, and intermediate results during inference.
Because FPGA accelerators repeatedly reuse the same localized memories and compute boundaries across addresses, tiles, modules, and layers, the attack can reuse a characterized physical structure as different assets traverse it.
We further showed that incomplete optical coverage does not necessarily prevent recovery.
Missing numerical weight entries can be completed from independently recovered input-output states when the corresponding linear system has sufficient rank, while a small number of unresolved bits can be recovered by propagating candidate completions to exact downstream states.
Our scalability analysis connects these mechanisms to the physical organization of real FPGA LLM accelerators and shows that imaging cost depends on both asset size and the amount of information recovered together at each observable boundary.

\cleardoublepage
\section*{Acknowledgments}
This effort was partially sponsored by NSF Grants CNS-2541809 and CNS-2150123, Longview Philanthropy, Hardware-Enabled Mechanisms (HEMs), as well as Survival \& Flourishing Fund (SFF-2024) Mechanisms for Flexible
Hardware-Enabled Guarantees (flexHEGs). 



\bibliographystyle{plainurl}
\bibliography{references}

\begin{thebibliography}{10}

\bibitem{achronix2024speedster}
{Achronix Semiconductor Corporation}.
\newblock Accelerating llm inferencing on fpgas, May 2024.
\newblock Published 9 May 2024; accessed 16 August 2026.
\newblock URL:
  \url{https://www.achronix.com/blog/accelerating-llm-inferencing-fpgas}.

\bibitem{amdUG574CLB}
{Advanced Micro Devices, Inc.}
\newblock {\em UltraScale Architecture Configurable Logic Block User Guide
  (UG574)}.
\newblock Advanced Micro Devices, Inc., 2025.
\newblock Revision 1.6.
\newblock URL: \url{https://docs.amd.com/r/en-US/ug574-ultrascale-clb}.

\bibitem{amdUG573Memory}
{Advanced Micro Devices, Inc.}
\newblock {\em UltraScale Architecture Memory Resources User Guide (UG573)}.
\newblock Advanced Micro Devices, Inc., 2025.
\newblock Revision 1.14, 18 November 2025.
\newblock URL:
  \url{https://docs.amd.com/r/en-US/ug573-ultrascale-memory-resources}.

\bibitem{Vivado:2023}
{Advanced Micro Devices, Inc.}
\newblock Vivado design suite.
\newblock \url{[Online]
  https://www.amd.com/en/products/software/adaptive-socs-and-fpgas/vivado.html}
  [Accessed: Aug. 6, 2026], 2026.

\bibitem{ainslie2023gqa}
Joshua Ainslie, James Lee-Thorp, Michiel de~Jong, Yury Zemlyanskiy, Federico
  Lebron, and Sumit Sanghai.
\newblock {GQA}: Training generalized multi-query transformer models from
  multi-head checkpoints.
\newblock In {\em Proceedings of the 2023 Conference on Empirical Methods in
  Natural Language Processing}, pages 4895--4901, Singapore, December 2023.
  Association for Computational Linguistics.
\newblock URL: \url{https://aclanthology.org/2023.emnlp-main.298/}, \href
  {https://doi.org/10.18653/v1/2023.emnlp-main.298}
  {\path{doi:10.18653/v1/2023.emnlp-main.298}}.

\bibitem{altera2023aiinference}
{Altera}.
\newblock Ai inference: Prefill-decode disaggregation.
\newblock Accessed 16 August 2026.
\newblock URL:
  \url{https://www.altera.com/fpga-solutions/datacenter/ai-inference}.

\bibitem{amini2024comparative}
Elham Amini, J{\"o}rg Jatzkowski, Tuba Kiyan, Lars Renkes, Thilo Krachenfels,
  Shahin Tajik, Christian Boit, Frank Altmann, Sebastian Brand, and Jean-Pierre
  Seifert.
\newblock Comparative study of e-beam and optical probing approaches in
  attacking the ics.
\newblock {\em Journal of Failure Analysis and Prevention}, 24(5):2184--2193,
  2024.

\bibitem{ba2016layernorm}
Jimmy~Lei Ba, Jamie~Ryan Kiros, and Geoffrey~E. Hinton.
\newblock Layer normalization.
\newblock In {\em arXiv preprint arXiv:1607.06450}, 2016.
\newblock URL: \url{https://arxiv.org/abs/1607.06450}.

\bibitem{souverana2026selfhosted}
Joel Barmettler.
\newblock {LLM selbst hosten: GPT-5-Niveau f{\"u}r 8,500 Franken}.
\newblock \url{https://souverana.ch/insights/llm-selbst-hosten/}, 2026.
\newblock Souverana, published 14 August 2026; accessed 24 August 2026.

\bibitem{beyreuther2020eofm}
A.~Beyreuther, N.~Herfurth, T.~Nakamura, G.~G. Fischer, S.~Keil, and C.~Boit.
\newblock Contactless device characterization of transistor structures in
  silicon using electro optical frequency mapping (eofm).
\newblock {\em Microelectronics Reliability}, 106:113583, 2020.
\newblock \href {https://doi.org/10.1016/j.microrel.2020.113583}
  {\path{doi:10.1016/j.microrel.2020.113583}}.

\bibitem{Buck008FPGA}
Buck008.
\newblock Transformer accelerator based on fpga, 2023.
\newblock URL:
  \url{https://github.com/Buck008/Transformer-Accelerator-Based-on-FPGA}.

\bibitem{chang2025hardware}
Yun-Nan Chang.
\newblock Hardware-software co-design for efficient llm inference on pcie-based
  fpgas using coarse-grained systolic arrays.
\newblock In {\em 2025 IEEE 38th International System-on-Chip Conference
  (SOCC)}, pages 1--5. IEEE, 2025.

\bibitem{chen2024spatialllm}
Hongzheng Chen, Jiahao Zhang, Yixiao Du, Shaojie Xiang, Zichao Yue, Niansong
  Zhang, Yaohui Cai, and Zhiru Zhang.
\newblock Understanding the potential of fpga-based spatial acceleration for
  large language model inference.
\newblock {\em ACM Transactions on Reconfigurable Technology and Systems},
  18(1):1--29, 2024.

\bibitem{chen2023high}
Yonghao Chen, Tianrui Li, Xiaojie Chen, Zhigang Cai, and Tao Su.
\newblock High-frequency systolic array-based transformer accelerator on field
  programmable gate arrays.
\newblock {\em Electronics}, 12(4):822, 2023.

\bibitem{Genesys2FPGA}
Digilent.
\newblock Genesys 2 reference manual, 2025.
\newblock URL:
  \url{https://digilent.com/reference/programmable-logic/genesys-2/reference-manual}.

\bibitem{elastixai2026fpga}
{ElastixAI}.
\newblock Five reasons why fpgas hit the sweet spot for llm inference.
\newblock
  \url{https://www.elastix.ai/blog/five-reasons-why-fpgas-hit-the-sweet-spot-for-llm-inference},
  February 2026.
\newblock Published 24 February 2026; accessed 16 August 2026.

\bibitem{gao2025localcache}
Zibo Gao, Junjie Hu, Feng Guo, Yixin Zhang, Yinglong Han, Siyuan Liu, Haiyang
  Li, and Zhiqiang Lv.
\newblock I know what you said: Unveiling hardware cache {Side-Channels} in
  local large language model inference.
\newblock In {\em 34th USENIX Security Symposium (USENIX Security 25)}, pages
  1649--1668, Seattle, WA, August 2025. USENIX Association.
\newblock URL:
  \url{https://www.usenix.org/conference/usenixsecurity25/presentation/gao-zibo}.

\bibitem{haris2024secdallm}
Jude Haris, Rappy Saha, Wenhao Hu, and Jos{\'e} Cano.
\newblock Designing efficient llm accelerators for edge devices.
\newblock {\em arXiv preprint arXiv:2408.00462}, 2024.

\bibitem{he2024hlstransform}
Andy He, Darren Key, Mason Bulling, Andrew Chang, Skyler Shapiro, and Everett
  Lee.
\newblock Hlstransform: Energy-efficient llama 2 inference on fpgas via high
  level synthesis.
\newblock {\em arXiv preprint arXiv:2405.00738}, 2024.

\bibitem{he2026lutllm}
Zifan He, Shengyu Ye, Rui Ma, Yang Wang, and Jason Cong.
\newblock Lut-llm: Efficient large language model inference with memory-based
  computations on fpgas.
\newblock {\em arXiv preprint arXiv:2511.06174}, 2025.

\bibitem{horvath2024physicalnn}
P{\'e}ter Horv{\'a}th, Dirk Lauret, Zhuoran Liu, and Lejla Batina.
\newblock {SoK}: Neural network extraction through physical side channels.
\newblock In {\em 33rd USENIX Security Symposium (USENIX Security 24)}, pages
  3403--3422, Philadelphia, PA, August 2024. USENIX Association.
\newblock URL:
  \url{https://www.usenix.org/conference/usenixsecurity24/presentation/horvath}.

\bibitem{huang2024edgellm}
Mingqiang Huang, Ao~Shen, Kai Li, Haoxiang Peng, Boyu Li, Yupeng Su, and Hao
  Yu.
\newblock Edgellm: A highly efficient cpu-fpga heterogeneous edge accelerator
  for large language models.
\newblock {\em IEEE Transactions on Circuits and Systems I: Regular Papers},
  72(7):3352--3365, 2025.

\bibitem{hur2023flexrun}
Suyeon Hur, Seongmin Na, Dongup Kwon, Joonsung Kim, Andrew Boutros, Eriko
  Nurvitadhi, and Jangwoo Kim.
\newblock A fast and flexible fpga-based accelerator for natural language
  processing neural networks.
\newblock {\em ACM Transactions on Architecture and Code Optimization},
  20(1):1--24, 2023.

\bibitem{bentoml2023llmhandbook}
Paul Iusztin and Maxime Labonne.
\newblock {\em LLM Engineer's Handbook}.
\newblock Packt Publishing, 2024.

\bibitem{li2025hummingbird}
Jindong Li, Tenglong Li, Ruiqi Chen, Guobin Shen, Dongcheng Zhao, Qian Zhang,
  and Yi~Zeng.
\newblock Hummingbird: A smaller and faster large language model accelerator on
  embedded fpga.
\newblock In {\em 2025 IEEE/ACM International Conference On Computer Aided
  Design (ICCAD)}, pages 1--9. IEEE, 2025.

\bibitem{liu2025pushinglimits}
Jindong Li, Tenglong Li, Guobin Shen, Dongcheng Zhao, Qian Zhang, and Yi~Zeng.
\newblock Pushing up to the limit of memory bandwidth and capacity utilization
  for efficient llm decoding on embedded fpga.
\newblock In {\em 2025 Design, Automation \& Test in Europe Conference (DATE)},
  pages 1--7. IEEE, 2025.

\bibitem{liu2022eofmvoltage}
Pengcheng Liu, Yingqi Ma, and Jianwei Han.
\newblock Preliminary study on detecting the internal voltage values of
  integrated circuits based on electro-optical frequency mapping.
\newblock {\em Applied Sciences}, 12(3):1188, 2022.
\newblock \href {https://doi.org/10.3390/app12031188}
  {\path{doi:10.3390/app12031188}}.

\bibitem{lohrke2016noplace}
Heiko Lohrke, Shahin Tajik, Christian Boit, and Jean-Pierre Seifert.
\newblock No place to hide: Contactless probing of secret data on fpgas.
\newblock In {\em Cryptographic Hardware and Embedded Systems -- CHES 2016},
  volume 9813 of {\em Lecture Notes in Computer Science}, pages 147--167.
  Springer, 2016.
\newblock \href {https://doi.org/10.1007/978-3-662-53140-2_8}
  {\path{doi:10.1007/978-3-662-53140-2_8}}.

\bibitem{luo2026shadowkv}
Zhifan Luo, Shuo Shao, Su~Zhang, Lijing Zhou, Yuke Hu, Chenxu Zhao, Zhihao Liu,
  and Zhan Qin.
\newblock Shadow in the cache: Unveiling and mitigating privacy risks of
  {KV-Cache} in {LLM} inference.
\newblock In {\em 33rd Annual Network and Distributed System Security Symposium
  (NDSS 2026)}. The Internet Society, 2026.
\newblock URL:
  \url{https://www.ndss-symposium.org/ndss-paper/shadow-in-the-cache-unveiling-and-mitigating-privacy-risks-of-kv-cache-in-llm-inference/},
  \href {https://doi.org/10.14722/ndss.2026.240258}
  {\path{doi:10.14722/ndss.2026.240258}}.

\bibitem{metaLlama2Repo}
{Meta AI}.
\newblock Llama 2 inference code and model configuration.
\newblock Official Meta Llama repository, 2023.
\newblock Accessed 16 August 2026.
\newblock URL: \url{https://github.com/meta-llama/llama}.

\bibitem{monfared2024laserescape}
Saleh~Khalaj Monfared, Kyle Mitard, Andrew Cannon, Domenic Forte, and Shahin
  Tajik.
\newblock Laserescape: Detecting and mitigating optical probing attacks.
\newblock In {\em Proceedings of the 43rd IEEE/ACM International Conference on
  Computer-Aided Design}, pages 224:1--224:10, 2024.
\newblock URL: \url{https://arxiv.org/abs/2405.03632}, \href
  {https://doi.org/10.1145/3676536.3676822}
  {\path{doi:10.1145/3676536.3676822}}.

\bibitem{nayan2024ondevice}
Tushar Nayan, Qiming Guo, Mohammed~Al Duniawi, Marcus Botacin, Selcuk Uluagac,
  and Ruimin Sun.
\newblock {SoK}: All you need to know about {On-Device} {ML} model extraction -
  the gap between research and practice.
\newblock In {\em 33rd USENIX Security Symposium (USENIX Security 24)}, pages
  5233--5250, Philadelphia, PA, August 2024. USENIX Association.
\newblock URL:
  \url{https://www.usenix.org/conference/usenixsecurity24/presentation/nayan}.

\bibitem{hamamatsu:2026}
Hamamatsu Photonics.
\newblock {PHEMOS X Microscope}.
\newblock \url{[Online]
  https://www.hamamatsu.com/eu/en/product/semiconductor-manufacturing-support-systems/failure-analysis-system/C15765-01.html}
  [Accessed: Aug. 6, 2026], 2026.

\bibitem{press2017outputembedding}
Ofir Press and Lior Wolf.
\newblock Using the output embedding to improve language models.
\newblock In {\em Proceedings of the 15th Conference of the European Chapter of
  the Association for Computational Linguistics: Volume 2, Short Papers}, pages
  157--163, Valencia, Spain, April 2017. Association for Computational
  Linguistics.
\newblock URL: \url{https://aclanthology.org/E17-2025/}.

\bibitem{saadfalcon2025intelligence}
Jon Saad-Falcon, Avanika Narayan, Hakki~Orhun Akengin, J~Griffin, Herumb
  Shandilya, Adrian~Gamarra Lafuente, Medhya Goel, Rebecca Joseph, Shlok
  Natarajan, Etash~Kumar Guha, et~al.
\newblock Intelligence per watt: Measuring intelligence efficiency of local ai.
\newblock {\em arXiv preprint arXiv:2511.07885}, 2025.

\bibitem{shazeer2020glu}
Noam Shazeer.
\newblock {GLU} variants improve transformer.
\newblock arXiv preprint arXiv:2002.05202, 2020.
\newblock URL: \url{https://arxiv.org/abs/2002.05202}.

\bibitem{su2021roformer}
Jianlin Su, Yu~Lu, Shengfeng Pan, Ahmed Murtadha, Bo~Wen, and Yunfeng Liu.
\newblock Roformer: Enhanced transformer with rotary position embedding.
\newblock {\em Neurocomputing}, 568:127063, 2024.
\newblock URL: \url{https://arxiv.org/abs/2104.09864}, \href
  {https://doi.org/10.1016/j.neucom.2023.127063}
  {\path{doi:10.1016/j.neucom.2023.127063}}.

\bibitem{tajik2017power}
Shahin Tajik, Heiko Lohrke, Jean-Pierre Seifert, and Christian Boit.
\newblock On the power of optical contactless probing: Attacking bitstream
  encryption of fpgas.
\newblock In {\em Proceedings of the 2017 ACM SIGSAC Conference on Computer and
  Communications Security}, pages 1661--1674, 2017.

\bibitem{touvron2023llama2}
Hugo Touvron, Louis Martin, Kevin Stone, Peter Albert, Amjad Almahairi, Yasmine
  Babaei, Nikolay Bashlykov, Soumya Batra, Prajjwal Bhargava, Shruti Bhosale,
  Dan Bikel, Lukas Blecher, Cristian~Canton Ferrer, Moya Chen, Guillem
  Cucurull, David Esiobu, Jude Fernandes, Jeremy Fu, Wenyin Fu, Brian Fuller,
  Cynthia Gao, Vedanuj Goswami, Naman Goyal, Anthony Hartshorn, Saghar
  Hosseini, Rui Hou, Hakan Inan, Marcin Kardas, Viktor Kerkez, Madian Khabsa,
  Isabel Kloumann, Artem Korenev, Punit~Singh Koura, Marie-Anne Lachaux,
  Thibaut Lavril, Jenya Lee, Diana Liskovich, Yinghai Lu, Yuning Mao, Xavier
  Martinet, Todor Mihaylov, Pushkar Mishra, Igor Molybog, Yixin Nie, Andrew
  Poulton, Jeremy Reizenstein, Rashi Rungta, Kalyan Saladi, Alan Schelten, Ruan
  Silva, Eric~Michael Smith, Ranjan Subramanian, Xiaoqing~Ellen Tan, Binh Tang,
  Ross Taylor, Adina Williams, Jian~Xiang Kuan, Puxin Xu, Zheng Yan, Iliyan
  Zarov, Yuchen Zhang, Angela Fan, Melanie Kambadur, Sharan Narang, Aurelien
  Rodriguez, Robert Stojnic, Sergey Edunov, and Thomas Scialom.
\newblock Llama 2: Open foundation and fine-tuned chat models.
\newblock arXiv preprint arXiv:2307.09288, 2023.
\newblock URL: \url{https://arxiv.org/abs/2307.09288}.

\bibitem{Vaswani2017Attention}
Ashish Vaswani, Noam Shazeer, Niki Parmar, Jakob Uszkoreit, Llion Jones,
  Aidan~N. Gomez, Lukasz Kaiser, and Illia Polosukhin.
\newblock Attention is all you need.
\newblock In {\em Advances in Neural Information Processing Systems},
  volume~30, 2017.
\newblock URL: \url{https://arxiv.org/abs/1706.03762}.

\bibitem{wang2025gameofarrows}
Pengli Wang, Bingyou Dong, Yifeng Cai, Zheng Zhang, Junlin Liu, Huanran Xue,
  Ye~Wu, Yao Zhang, and Ziqi Zhang.
\newblock Game of arrows: On the ({In-)Security} of weight obfuscation for
  {On-Device} {TEE-Shielded} {LLM} partition algorithms.
\newblock In {\em 34th USENIX Security Symposium (USENIX Security 25)}, pages
  279--298, Seattle, WA, August 2025. USENIX Association.
\newblock URL:
  \url{https://www.usenix.org/conference/usenixsecurity25/presentation/wang-pengli}.

\bibitem{wojtal2024eofmmitigation}
Thomas Wojtal, Robi Paul, and Michael Zuzak.
\newblock Mitigating electro-optical frequency mapping attacks on logic-locked
  integrated circuits.
\newblock {\em Journal of Hardware and Systems Security}, 8:233--243, 2024.
\newblock Published online 29 January 2025.
\newblock \href {https://doi.org/10.1007/s41635-025-00158-w}
  {\path{doi:10.1007/s41635-025-00158-w}}.

\bibitem{xu2024llamaf}
Han Xu, Yutong Li, and Shihao Ji.
\newblock Llamaf: An efficient llama2 architecture accelerator on embedded
  fpgas.
\newblock In {\em 2024 IEEE 10th World Forum on Internet of Things (WF-IoT)},
  pages 1--7. IEEE, 2024.

\bibitem{zeng2024flightllm}
Shulin Zeng, Jun Liu, Guohao Dai, Xinhao Yang, Tianyu Fu, Hongyi Wang, Wenheng
  Ma, Hanbo Sun, Shiyao Li, Zixiao Huang, et~al.
\newblock Flightllm: Efficient large language model inference with a complete
  mapping flow on fpgas.
\newblock In {\em Proceedings of the 2024 ACM/SIGDA International Symposium on
  Field Programmable Gate Arrays}, pages 223--234, 2024.

\bibitem{zhang2019rmsnorm}
Biao Zhang and Rico Sennrich.
\newblock Root mean square layer normalization.
\newblock In {\em Advances in Neural Information Processing Systems},
  volume~32, pages 12360--12371, 2019.
\newblock URL: \url{https://arxiv.org/abs/1910.07467}.

\bibitem{zhang2019fault}
Jeff~Jun Zhang, Kanad Basu, and Siddharth Garg.
\newblock Fault-tolerant systolic array based accelerators for deep neural
  network execution.
\newblock {\em IEEE Des. Test}, 36(5):44--53, 2019.

\bibitem{zuzak2022clap}
Michael Zuzak, Yuntao Liu, Isaac McDaniel, and Ankur Srivastava.
\newblock A combined logical and physical attack on logic obfuscation.
\newblock In {\em 2022 IEEE/ACM International Conference on Computer-Aided
  Design (ICCAD)}, pages 1--8, 2022.
\newblock \href {https://doi.org/10.1145/3508352.3549349}
  {\path{doi:10.1145/3508352.3549349}}.

\end{thebibliography}

\end{document}